\documentclass[5p,times,twocolumn,final]{elsarticle}

\usepackage{framed,multirow}
\usepackage{placeins}
\usepackage{stfloats}
\usepackage{amssymb}
\usepackage{latexsym}

\usepackage{url}
\usepackage[table,xcdraw]{xcolor}
\definecolor{bibblue}{HTML}{00467E}
\usepackage{hyperref}
\usepackage{orcidlink}
\makeatletter
\newcommand{\authorwithorcid}[3]{%
  \author[#1]{#2}%
  \g@addto@macro\elsauthors{\hspace{0.12em}\orcidlink{#3}}%
}
\makeatother

\hypersetup{
  colorlinks=true,
  urlcolor=bibblue
}
\AtBeginDocument{}

\definecolor{newcolor}{rgb}{.8,.349,.1}

\definecolor{dbBlue}{RGB}{65,125,175}
\definecolor{dbRed}{RGB}{190,95,90}
\definecolor{dbGray}{RGB}{120,120,120}

\usepackage{afterpage}
\usepackage{changepage}
\usepackage{url}
\usepackage{hyperref}
\usepackage[version=4]{mhchem}

\usepackage{ulem}

\usepackage{amssymb}
\usepackage{amsthm}
\usepackage{amsmath}
\usepackage{bibentry}
\biboptions{sort&compress}

\usepackage{siunitx}
\usepackage{enumitem}

\usepackage[german,english]{babel}

\usepackage{transparent}
\usepackage{csvsimple}
\usepackage{pict2e}
\usepackage{overpic}
\usepackage{multirow}
\usepackage{booktabs, tabularx, array}
\usepackage{printlen}
\uselengthunit{in}

\begin{document}

\begin{frontmatter}

\title{{\fontsize{17pt}{20pt}\selectfont\bfseries A Systematic Assessment of Artificial Flame Thickening for Thermodiffusively Unstable Lean Hydrogen Flames in Homogeneous Isotropic Turbulence}}

\authorwithorcid{1}{Max Schneider\corref{cor1}}{0000-0001-5166-6273}
\authorwithorcid{1}{Vinzenz Schuh}{0009-0003-8706-1008}
\authorwithorcid{1}{Felix Rong}{0000-0002-7837-0939}
\authorwithorcid{1}{Driss Kaddar}{0000-0002-7812-4738}
\authorwithorcid{2}{Andrew J. Aspden}{0000-0002-2970-4824}
\authorwithorcid{3}{Thomas L. Howarth}{0000-0002-7789-9742}
\authorwithorcid{1}{Christian Hasse}{0000-0001-9333-0911}
\authorwithorcid{4}{Hendrik Nicolai}{0000-0002-0355-2252}
\cortext[cor1]{Corresponding author}
\emailauthor{schneider@stfs.tu-darmstadt.de}{Max Schneider}

\address[1]{Technical University of Darmstadt, Department of Mechanical Engineering, Simulation of Reactive Thermo-Fluid Systems, Otto-Berndt-Str.~2, 64287 Darmstadt, Germany}
\address[2]{School of Engineering, Newcastle University, Newcastle-Upon-Tyne, NE1 7RU, UK}
\address[3]{Department of Aeronautical and Automotive Engineering, Loughborough University, Loughborough, LE11 3TU, UK}
\address[4]{McGill University, Department of Mechanical Engineering, 817 Sherbrooke Street West, H3A 0C3, Montreal, Canada}
\makeatletter
\let\elsauthorsOriginal\elsauthors
\def\elsauthors{{\fontsize{13pt}{15pt}\selectfont\elsauthorsOriginal}}
\makeatother

\begin{abstract}
The Artificially Thickened Flame (ATF) approach is widely used to render premixed flames resolvable on coarse computational meshes, but thickening alters their interaction with turbulence.
While considerable research has focused on the effects of thickening on turbulence-induced flame wrinkling, its influence on the mean local flame propagation speed in turbulent, thermodiffusively unstable hydrogen flames remains poorly understood.
More than 200 flame-resolved simulations of premixed flames in homogeneous isotropic turbulence with detailed chemistry and transport are used to systematically isolate and quantify the effects of artificial thickening.

Thickening the flame by a factor $F$ increases its effective flame-scale Karlovitz number according to $Ka_{\mathrm{f}}^{F}=Ka_{\mathrm{f}}\sqrt{F}$, while decreasing its effective Damköhler number according to $Da_{\mathrm{f}}^{F}=Da_{\mathrm{f}}/F$.
In the regime considered here, the former enhances the mean local flame speed, whereas the latter reduces flame wrinkling, resulting in opposing effects on the flame consumption speed; hence, changes in both local flame propagation and flame area must be accounted for. 
Simulations matched with respect to $Ka_{\mathrm{f}}^{F}$ and $Da_{\mathrm{f}}^{F}$ collapse across different thickening factors $F$ until the thickened flame approaches the integral or geometric length scales of the configuration.
Based on these scaling relations, a self-consistent efficiency correction is proposed to recover the unthickened flame consumption speed.
Since the efficiency factor feeds back on the effective Karlovitz and Damköhler numbers, it must be determined implicitly.
\end{abstract}


\end{frontmatter}

\section{Introduction}
\label{sec:introduction}

Hydrogen (\ce{H2}) combustion offers a promising route toward sustainable, carbon-free power and propulsion and can mitigate adverse effects associated with fossil-fuel combustion, such as greenhouse-gas emissions and soot formation~\cite{dreizler2021}.
In premixed \ce{H2} combustion, lean operation is particularly attractive because the reduced adiabatic flame temperature limits thermal \ce{NO_x} formation, while lower laminar flame speeds potentially reduce the susceptibility to flame flashback \cite{gruhlke2026}.

However, lean \ce{H2}--air flames are strongly affected by thermodiffusive (TD) instabilities, primarily due to the low Lewis number of \ce{H2} and the associated differential- and preferential-diffusion effects.
The resulting flames exhibit cellular patterns and, on average, burn faster and hotter, with smaller characteristic flame thicknesses than their corresponding one-dimensional unstretched counterparts~\cite{kadowaki2005,altantzis2011,altantzis2012,frouzakis2015,berger2019,creta2020,howarth2022,howarth2023,schneider2025a,schneider2025b}.

Moreover, TD effects interact strongly with turbulence and remain pronounced under turbulent conditions~\cite{im2002,bell2007,day2009,aspden2011b,berger2022c,howarth2023,yao2024a,wang2024,lapenna2025}.
Consequently, accurately predicting the flame propagation and consumption speeds of turbulent lean \ce{H2} flames is considerably more challenging than for flames with Lewis numbers closer to unity.

Nevertheless, accurately predicting these flame characteristics under turbulent conditions is crucial, as practical combustion systems typically operate in such regimes.

Direct numerical simulation (DNS) provides a suitable reference for studying the underlying flame--turbulence interactions without turbulence or combustion closure models.
Its computational cost, however, precludes application to practical combustors, motivating the use of large-eddy simulation (LES).

In LES, the flame structure is generally not resolved on the grid, and an additional turbulence--chemistry interaction closure is required~\cite{pitsch2006}.
A widely used approach is the artificially thickened flame (ATF) model~\cite{butler1977,oRourke1979}.
The ATF approach increases the flame thickness by a factor $F$, allowing the flame structure to be resolved on the LES grid.
This is achieved by multiplying molecular diffusion by $F$ and dividing the chemical source terms by $F$, such that the laminar flame speed remains unchanged.
A key advantage of ATF is its independence from the employed chemistry modelling approach.
While this approach preserves the propagation of an unconfined unstretched flame, thickening modifies its interaction with external length scales imposed by the flow or geometry.

To account for the reduced flame wrinkling caused by the modified flame--turbulence interaction of an artificially thickened flame, Colin et al.~\cite{colin2000} introduced an efficiency function $E$, which was subsequently reformulated by Charlette et al.~\cite{charlette2002}; both formulations were calibrated using databases of two-dimensional flame--vortex interactions.
$E$ enhances the effective flame speed by additionally multiplying the diffusion term and the chemical source term by $E$, while leaving the prescribed flame thickness unchanged.
They have predominantly been applied to hydrocarbon flames~\cite{selle2004,schmitt2007,durand2007,wang2011,kuenne2011,kuenne2012,nicolai2022}, achieving promising results.
For \ce{H2} flames, however, the interplay between TD effects and artificial flame thickening under turbulent conditions, including their impact on local reactivity and flame wrinkling, remains insufficiently understood \cite{traut2026a}.

Schuh et al.~\cite{schuh2024} developed a TD-instability-aware extension of the ATF model for two-dimensional laminar \ce{H2} flames, introducing a fractal-like $E$ that compensates for thickening-induced changes in instability length scales and recovers the flame consumption speed.
Their results showed that flame thickening remains fundamentally a length-scale problem also for thermodiffusively unstable flames: both flame wrinkling and the enhancement of the mean local flame speed remain partially resolved and scale with the ratio of the available geometric length scale to the thickened flame thickness.
Subsequent work~\cite{schuh2025} generalized the TD efficiency-function model to \ce{H2}, \ce{H2}/\ce{CH4}, and \ce{H2}/\ce{NH3} flames across a range of operating conditions and combined it with the flame-wrinkling efficiency function of Charlette et al.~\cite{charlette2002} for application to turbulent flames.
Schneider et al.~\cite{schneider2026c} further extended the framework using three-dimensional simulations of laminar TD unstable \ce{H2} flames over a range of domain sizes, coupled it with a flamelet-based combustion model, and applied it to turbulent \ce{H2} jet flames, where promising agreement was obtained.

While the effects of artificial thickening on TD unstable laminar flames are comparatively well understood~\cite{schuh2024}, its interaction with turbulence has largely been described through specific modelling assumptions \cite{traut2026a,schuh2025,schneider2026c}.
The present study therefore establishes a systematic physical understanding of artificial flame thickening in turbulent TD unstable \ce{H2} flames.

To this end, flame-in-a-box (FIAB) simulations in homogeneous isotropic turbulence (HIT) are considered.
This canonical configuration has been extensively employed to study flame--turbulence interactions in hydrocarbon flames~\cite{savard2015,aspden2016,aspden2017b,chaudhuri2023} and, more recently, in TD unstable \ce{H2} flames~\cite{day2009,aspden2011b,aspden2011c,aspden2017a,howarth2023,yao2024a,hunt2025}.

More than 200 three-dimensional simulations of lean premixed hydrogen--air flames are performed using detailed reaction kinetics and mixture-averaged transport, systematically varying the Karlovitz and Damköhler numbers, the integral and geometric length scales, and the thickening factor $F$.
No efficiency function is applied, such that the direct effects of artificial thickening on both local flame propagation and flame wrinkling remain exposed.

Specifically, the objectives are to:
\begin{itemize}[leftmargin=*,nosep]
\item quantify how flame thickening modifies the Karlovitz and Damköhler numbers and thereby affects small- and large-scale turbulence--chemistry interactions, as reflected in the mean local flame speed and flame wrinkling;
\item determine how the interaction between the thickened flame and external geometric or flow length scales changes as the flame thickness approaches these scales, and clarify the resulting effects on local flame propagation and flame wrinkling across different turbulent operating conditions;
\item use the systematic trends identified from the simulation database to derive a physically motivated pathway for future ATF modelling of lean \ce{H2} flames, in particular by correcting the thickening-induced changes in Karlovitz and Damköhler numbers with the ultimate objective of recovering the turbulent consumption speed in ATF LES of TD unstable hydrogen flames.
\end{itemize}

\section{Theoretical background}
\label{sec:theory}

This section summarises the relevant turbulence--flame interaction parameters and TD response of lean \ce{H2} flames before deriving how artificial flame thickening modifies the characteristic flame scales and, consequently, the governing nondimensional parameters.
Particular attention is given to the implications for local flame propagation, flame wrinkling, and scale separation.

\paragraph{Turbulence--flame interaction parameters}
In premixed turbulent flames, turbulence--flame interaction can be characterised through the normalised turbulent velocity and integral length scales,
\begin{equation}
    \Upsilon_r = \frac{u'}{s_r} \, ,
    \qquad
    \Lambda_r = \frac{L_{\mathrm{int}}}{l_r} \, ,
\end{equation}
or, equivalently, through the Karlovitz and Damköhler numbers,
\begin{equation}
    Ka_r
    =
    \sqrt{
        \left(\frac{u'}{s_r}\right)^3
        \frac{l_r}{L_{\mathrm{int}}}
    }
    =
    \sqrt{\frac{\Upsilon_r^3}{\Lambda_r}} \, ,
    \,
    Da_r
    =
    \frac{L_{\mathrm{int}}}{l_r}
    \frac{s_r}{u'}
    =
    \frac{\Lambda_r}{\Upsilon_r} \, .
    \label{eq:Ka_Da}
\end{equation}
Here, the subscript $r$ denotes the reference flame used to define the characteristic flame speed $s_r$ and thickness $l_r$.
\footnote{The subscript $r$ refers to either the one-dimensional unstretched laminar flame ($r=\mathrm{l}$) or the three-dimensional freely propagating TD unstable flame ($r=\mathrm{f}$).}
The Damköhler number, $Da_r=\tau_{\mathrm{int}}/\tau_r$ with $\tau_{\mathrm{int}}=L_{\mathrm{int}}/u'$ and $\tau_r=l_r/s_r$, compares the integral turbulent and characteristic flame time scales and characterises large-scale turbulence--flame interaction.
Using the turbulent kinetic-energy dissipation rate $\varepsilon=(u')^3/L_{\mathrm{int}}$, the Karlovitz number can equivalently be written as $Ka_r^2=\varepsilon l_r/s_r^3$ \cite{aspden2019}.
Since $\varepsilon$ is approximately invariant within the inertial subrange, this formulation relates $Ka_r$ to turbulence at the flame scale independently of the larger turbulent length scales~\cite{aspden2011b,aspden2019,hunt2025}.
This flame-scale definition of the Karlovitz number differs from the conventional definition based on Kolmogorov scales; its physical motivation and relation to the conventional definition are discussed in \cite{aspden2019}.

To relate these turbulence--flame interaction parameters to the resulting flame propagation, the flame consumption speed $s_{\mathrm{c}}$ can be decomposed into contributions from the mean local flame speed and flame wrinkling,
\begin{equation}
    \frac{s_{\mathrm{c}}}{s_{\mathrm{l}}}
    =
    I_0\Psi \, ,
    \quad
    I_0=\frac{s_{\mathrm{s}}}{s_{\mathrm{l}}} \, ,
    \quad
    \Psi=\frac{A}{A_0}\!=\!\frac{s_{\mathrm{c}}}{s_{\mathrm{s}}} \, ,
    \label{eq:sc_I0_Psi}
\end{equation}
where $s_{\mathrm{s}}$ denotes the surface-averaged local flame speed, $A$ the instantaneous flame-surface area, and $A_0$ the area of the temporally averaged flame surface.
Thus, $I_0$ characterises modifications of the mean local flame speed, whereas $\Psi$ quantifies flame wrinkling.

\paragraph{Thermodiffusive response and flame wrinkling in lean \ce{H2} flames}

As discussed in the Introduction, TD effects enhance the mean local flame propagation speed and reduce the mean local flame thickness of lean \ce{H2} flames. Multidimensional freely propagating laminar flames allow these intrinsic TD effects to be characterised and quantified independently of turbulence, thereby providing a reference for quantifying additional turbulence-induced modifications.
Howarth et al.~\cite{howarth2022} developed models for the mean local flame speed and flame thickness from two-dimensional TD-unstable freely propagating flames using only quantities obtained from one-dimensional laminar flame calculations.
These models were subsequently refitted using three-dimensional laminar flames and extended to turbulent conditions by Howarth et al.~\cite{howarth2023}.

For low-to-moderate Karlovitz numbers, their simulations showed that the turbulence-induced enhancement of the mean local flame speed is primarily governed by the flame-scale Karlovitz number.
Based on these findings, Howarth et al.~\cite{howarth2023} proposed an empirical correlation for the surface-averaged local flame speed over $Ka_{\mathrm{f}}\approx1$--$36$, that was shown to provide a good description over a broad range of operating conditions~\cite{gauding2026},
\begin{equation}
    \frac{s_{\mathrm{s}}}{s_{\mathrm{f}}}
    =
    1
    +
    0.26\exp(-0.038\omega_2)\sqrt{Ka_{\mathrm{f}}},
    \label{eq:ss_sf_fit_Kaf}
\end{equation}
where $\omega_2$ characterises the strength of the TD instability~\cite{matalon2003,howarth2022,howarth2023}.

Hunt and Aspden~\cite{hunt2025} subsequently varied the integral length scale, and hence the Damköhler number, independently and found both the mean local flame speed and the distribution of local flame speeds to be approximately independent of $Da_{\mathrm{f}}$ over the investigated range.
Similar observations were reported by Yao and Blanquart~\cite{yao2024a}, who additionally observed a non-monotonic TD response with increasing Karlovitz number at sufficiently high $Ka$.
Together, these results indicate that the mean local burning response is primarily governed by turbulence at the flame scale, whereas the integral turbulent scales mainly control flame wrinkling.

In contrast to the local burning response, flame wrinkling depends on both the turbulent velocity and integral length scales.
The two classical limits introduced by Damköhler~\cite{damkoehler1940,peters1999,peters2000} give $\Psi-1\sim\Upsilon_r$ in the large-scale limit (LSL) and $\Psi-1\sim\Upsilon_r Da_r^{1/2}$ in the small-scale limit (SSL).
Peters~\cite{peters2000} proposed a G-equation-based interpolation that continuously connects these two limits, such that
\begin{equation}
\frac{\Psi-1}{\Upsilon_r}
=
\begin{cases}
    \gamma_r Da_r^{1/2}, & \mathrm{SSL} \, , \\
    \alpha_r, & \mathrm{LSL} \, , \\
    2\alpha_r\left(\sqrt{\xi_r^2+\xi_r}-\xi_r\right),
        & \mathrm{Peters} \, ,
\end{cases}
\label{eq:ssl_lsl_peters}
\end{equation}
with $\xi_r=Da_r/\beta_r$ and $\beta_r=4\alpha_r^2/\gamma_r^2$.

For lean \ce{H2} flames in FIAB configurations, Hunt and Aspden~\cite{hunt2025} showed that the SSL scaling accurately describes flame wrinkling under the conditions they considered.
At matched turbulent conditions, they further found the wrinkling of unity- and non-unity-Lewis-number flames with $Le<1$ to be similar, with only a moderate increase in wrinkling as the Lewis number decreases.

This supports the interpretation originally proposed by Peters~\cite{peters2000} that established wrinkling models may also be applied to non-unity-Lewis-number flames if the reference flame speed is replaced by an appropriate mean local flame speed that accounts for stretch effects.
Consistent with this concept, Hunt and Aspden~\cite{hunt2025} showed that the turbulent consumption speed of their lean \ce{H2} flames can be accurately predicted by combining models for the mean local flame speed and flame wrinkling.

\paragraph{Artificial flame thickening and scale separation}

The ATF approach increases the flame thickness by a factor $F$ by multiplying molecular diffusion by $F$ and dividing the chemical source terms by $F$~\cite{butler1977,oRourke1979}.
With the characteristic scalings $l_r\sim\sqrt{D/\dot{\omega}}$ and $s_r\sim\sqrt{D\dot{\omega}}$, the resulting flame scales are
\begin{equation}
    l_r^{F}=F l_r\,,\;
    s_r^{F}=s_r\,,\;
    \tau_r^{F}=F\tau_r\,,
\end{equation}
where $D$ is a characteristic molecular diffusivity and $\dot{\omega}$ denotes a characteristic chemical source term.
\footnote{Throughout, the superscript $F$ denotes effective quantities associated with the artificially thickened flame represented in the simulation, whereas quantities without this superscript refer to the corresponding unthickened target flame.}
Since thickening does not actively modify the flow field, the turbulent velocity and integral length scales remain unchanged, which leads to
\begin{equation}
    \Upsilon_r^{F}=\Upsilon_r\,,\;
    \Lambda_r^{F}=\frac{\Lambda_r}{F}\,,
\end{equation}
and consequently
\begin{equation}
    Ka_r^{F}=\sqrt{F}\,Ka_r\,,\;
    Da_r^{F}=\frac{Da_r}{F}\,.
    \label{eq:KaF_DaF}
\end{equation}
Artificial thickening therefore simultaneously increases the effective flame-scale Karlovitz number and decreases the effective Damköhler number.
For TD unstable \ce{H2} flames, these two changes act on the two contributions to the consumption speed in opposite ways.
Based on Eq.~\eqref{eq:ss_sf_fit_Kaf}, the increase in $Ka_{\mathrm{f}}^{F}$ is expected to enhance the mean local flame speed within the regime considered here, whereas the decrease in $Da_{\mathrm{f}}^{F}$ reduces flame wrinkling.

In contrast to the wrinkling models discussed above, which relate the total flame wrinkling to characteristic turbulent velocity and length scales, LES-specific formulations have been developed to account for the unresolved wrinkling associated with the altered effective Damköhler number.
These approaches are commonly referred to as subgrid-scale flame-wrinkling models~\cite{kuenne2012_phd}.
Based on these models, conventional efficiency-function formulations~\cite{colin2000,charlette2002} compensate for the thickening-induced modification of flame wrinkling by defining an efficiency factor $E$ from the ratio of the wrinkling factors of the unthickened and thickened flames.
The factor $E$ is applied to both the molecular diffusion and chemical source terms, thereby increasing the effective flame propagation speed to recover the corresponding turbulent consumption speed.

Besides modifying $Ka_r$ and $Da_r$, artificial flame thickening also changes the relative separation between the characteristic flame, turbulent, and geometric length scales.
Geometric constraints associated with this scale separation have previously been discussed for laminar flames by Schuh et al.~\cite{schuh2024}.

\begin{figure}[t]
    \centering
    \includegraphics{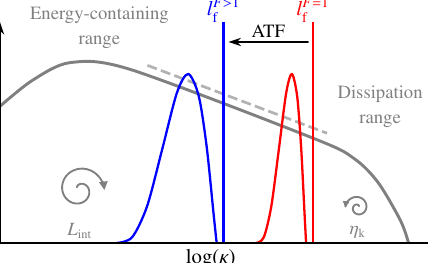}
    \caption{Schematic turbulent energy spectrum and distribution of characteristic flame scales in wavenumber space.
    Artificial thickening shifts the characteristic flame scales toward smaller wavenumbers, corresponding to larger physical length scales, while the turbulent spectrum is not actively modified by the thickening procedure.
    With increasing $F$, the thickened flame scale therefore approaches or exceeds the integral length scale.}
    \label{fig:energy_spectra}
\end{figure}
The corresponding change in turbulence--flame scale separation is illustrated schematically in Fig.~\ref{fig:energy_spectra}.
While the turbulent spectrum remains unchanged, artificial thickening shifts the characteristic flame scales toward larger physical length scales.
In particular,
\begin{equation}
    \Lambda_{\mathrm{f}}^{F} = {L_{\mathrm{int}}}/{l_{\mathrm{f}}^{F}} = Ka_{\mathrm{f}}^{F} \left(Da_{\mathrm{f}}^{F}\right)^{3/2}.
    \label{eq:LambdaF_KaF_DaF}
\end{equation}
Hence, $\Lambda_{\mathrm{f}}^{F}\approx1$ marks the regime in which the thickened flame thickness becomes comparable to the integral turbulent length scale.
As this limit is approached, the interaction with the energy-containing turbulent scales may therefore become increasingly important.

The relevance of the regime $\Lambda_{\mathrm{f}}^{F}\lesssim1$ depends on the turbulence-modelling framework, namely LES or Reynolds-averaged Navier--Stokes (RANS) simulations.
In LES, which is intended to resolve the energy-containing part of the turbulent spectrum, this scale separation can additionally be related to the numerical resolution requirements of the turbulent flow and the flame.
A commonly used guideline is that approximately $80\%$ of the turbulent kinetic energy should be resolved~\cite{pope2000}.
Assuming a Kolmogorov inertial-range spectrum $E(\kappa)=C_K\varepsilon^{2/3}\kappa^{-5/3}$, where $\kappa$ denotes the wavenumber, $C_K$ is the Kolmogorov constant, and $\varepsilon$ is the turbulent kinetic energy dissipation rate, the resolved turbulent kinetic energy fraction can be estimated by integrating the spectrum from the characteristic energy-containing wavenumber $\kappa_{\mathrm{int}}\sim1/L_{\mathrm{int}}$ to the grid-cutoff wavenumber $\kappa_{\Delta}\sim1/\Delta_x$, and normalising by the integral from $\kappa_{\mathrm{int}}$ to infinity.
This yields $k_{\mathrm{res}}/k\approx\left(\int_{\kappa_{\mathrm{int}}}^{\kappa_\Delta}E(\kappa)\,\mathrm{d}\kappa\right)/\left(\int_{\kappa_{\mathrm{int}}}^{\infty}E(\kappa)\,\mathrm{d}\kappa\right)\approx1-\left(\Delta_x/L_{\mathrm{int}}\right)^{2/3}$, where $k_{\mathrm{res}}$ and $k$ denote the resolved and total turbulent kinetic energies, respectively, and $\Delta_x$ is the grid spacing.
The prefactor $C_K\varepsilon^{2/3}$ cancels upon normalisation.
Within this simplified spectral approximation, $80\%$ resolved turbulent kinetic energy corresponds to $L_{\mathrm{int}}/\Delta_x\approx11$--$12$.
At the same time, approximately 12 grid points across the flame thickness $l_{\mathrm{f}}^{F}$ are required to adequately resolve the flame structure~\cite{howarth2022}.
If the flame is therefore thickened only to the extent required for its numerical resolution, $l_{\mathrm{f}}^{F}/\Delta_x\approx12$, these two requirements meet at $\Lambda_{\mathrm{f}}^{F}\approx1$.
Consequently, excessive flame thickening beyond this point is undesirable: the artificial flame thickness then approaches or exceeds the energy-containing turbulent scales that an LES is intended to resolve.

In contrast, in RANS and unsteady RANS (URANS) simulations, no analogous grid-resolution criterion applies because the turbulent scales are modelled rather than explicitly resolved.
Nevertheless, $\Lambda_{\mathrm{f}}^{F}\lesssim1$ still represents a fundamental change in scale separation, for which the thickened flame becomes comparable to or larger than the integral turbulent scale.
The present study therefore explicitly investigates whether the scaling with the effective Karlovitz and Damköhler numbers remains valid for $\Lambda_{\mathrm{f}}^{F}\leq1$.

Overall, artificial flame thickening therefore introduces a complex interplay between turbulent, flame, and geometric length and time scales that has so far been neither systematically analysed nor fully understood, particularly for TD unstable flames.

\section{Numerical setup and configuration}
\label{sec:num_setup}

The simulations were performed using the adaptive-mesh-refinement solver \textsc{PeleLMeX}~\cite{day2000,esclapez2023}, which solves the low-Mach-number Navier--Stokes equations for reacting flows.
Further details on the numerical formulation and its application to premixed hydrogen flames can be found, for example, in~\cite{howarth2024,schneider2026b,hunt2025}.

The fluid is treated as a mixture of ideal gases, and molecular transport is modelled using a mixture-averaged diffusion formulation.
Dufour, Soret, gravitational, and radiative effects are neglected.
Although Soret diffusion is known to enhance TD effects in hydrogen flames~\cite{grcar2009}, it does not constitute a leading-order contribution under the conditions considered here and is therefore not expected to alter the phenomenology or conclusions of the present study~\cite{howarth2024}.
Chemical kinetics, transport properties, and thermodynamic data are taken from~\cite{burke2011}.

The simulations employ a canonical FIAB configuration, similar to that used in previous studies~\cite{aspden2011c}.
HIT is generated and maintained by a forcing term in the momentum equations following the approach described in~\cite{aspden2009}.
A high-aspect-ratio computational domain of size $L_x\times L_y\times L_z$ is used, with $L=L_x=L_y$ and $L_z\geq 4L$, such that the flame propagates through statistically homogeneous turbulent conditions over an extended distance.
The flame is initialized near the upper end of the domain using a one-dimensional, unstretched laminar flame profile obtained with \textsc{Cantera}~\cite{goodwin2024}, resulting in downward flame propagation.
Periodic boundary conditions are applied in the lateral directions, an outflow condition is prescribed at the upper boundary, and the lower boundary is treated as adiabatic.

A characteristic property of the forcing strategy is the nearly constant ratio
$\Omega=L_{\mathrm{domain}}/L_{\mathrm{int}}\approx 10$,
where $L_{\mathrm{int}}$ denotes the turbulent integral length scale.
For selected cases employing periodic replication of the forcing field in the lateral directions, the domain size is instead given by
$L_{\mathrm{domain}}=M\,10L_{\mathrm{int}}$, with $M\in\mathbb{N}$, as discussed below.

Using adaptive mesh refinement (AMR), the finest mesh resolution is chosen such that the flame thickness $l_{\mathrm{f}}$ is resolved by $N_{\mathrm{f}}=12$ grid cells, consistent with~\cite{howarth2023}.
The corresponding effective number of cells in each lateral direction at the finest refinement level is
$
N
=
{L}/{\Delta x}
=
\left({l_{\mathrm{f}}}/{\Delta x}\right)
\left({L_{\mathrm{int}}}/{l_{\mathrm{f}}}\right)
\left({L_{\mathrm{domain}}}/{L_{\mathrm{int}}}\right)
=
N_{\mathrm{f}}\Lambda_{\mathrm{f}}\Omega.
$
All simulations employ a base grid and one level of AMR; further details on the AMR strategy are provided in~\cite{aspden2011c,schneider2026b}.

For simulations employing artificial flame thickening ($F>1$), the species diffusion coefficients $D_i$ and the thermal conductivity $\lambda$ appearing in the species and enthalpy transport equations are multiplied by the thickening factor $F$, as described in Sec.~\ref{sec:theory}.
A spatially constant $F$ is applied throughout the computational domain.
No flame sensor is employed to ensure spatially uniform model parameters, allowing their effects on flame propagation to be investigated under well-defined conditions without additional spatial variations (here, variation in $F$).
This approach is justified for the present fully premixed configurations without mixture stratification or wall heat losses, where species and enthalpy gradients are negligible outside the flame region.
Consequently, artificial thickening does not significantly alter diffusive transport in the unburned and burned gases.

For all thickened-flame simulations, the resulting flame thickness $Fl_{\mathrm{f}}$ is also resolved by at least $N_{\mathrm{f}}=12$ grid cells.
Turbulent motions at scales comparable to the flame thickness remain sufficiently resolved, whereas substantially smaller scales have only a minor influence on flame wrinkling and stretching under the conditions considered here, consistent with the discussion in~\cite{aspden2019}.
For selected thickened-flame conditions, additional simulations were performed at increased spatial resolution (using an additional AMR level).
No significant differences were observed in the relevant flame quantities, including the global flame consumption speed and local flame speeds; corresponding resolution comparisons are provided in the Supplementary Material (Fig.~S.1).

All simulations are performed at a single TD unstable operating condition, $\phi=0.4$, $T_{\mathrm{u}}=300~\mathrm{K}$, and $p=1~\mathrm{atm}$, for which $\omega_2=5.45$~\cite{howarth2023}.

\begin{table*}[t]
\centering
\caption{Overview of the simulation databases, including their respective parameter ranges, fixed and varied parameters, and objectives.
Fixed quantities are held constant within each parameter sweep.
}
\label{tab:databases}

\footnotesize
\setlength{\tabcolsep}{3.5pt}
\renewcommand{\arraystretch}{1.2}

\begin{tabularx}{\textwidth}{
    @{}l
    >{\raggedright\arraybackslash}p{0.205\textwidth}
    >{\raggedright\arraybackslash}p{0.14\textwidth}
    >{\raggedright\arraybackslash}p{0.16\textwidth}
    >{\raggedright\arraybackslash}X@{}
}
\toprule
Database & Parameter range & Fixed & Varied & Objective \\
\midrule

\textit{A}
& $Ka_{\mathrm{f}}^{F}\in\{1,4,12,36\}$,
  $\Lambda_{\mathrm{f}}^{F}=2$,
  $F\in\{1,2,4,8,16,32\}$
& $Ka_{\mathrm{f}}^{F}$,
  $\Lambda_{\mathrm{f}}^{F}$
& $F$,
  $Ka_{\mathrm{f}}=Ka_{\mathrm{f}}^{F}/\sqrt{F}$,
  $\Lambda_{\mathrm{f}}=2F$
& Assess flame similarity at identical effective dimensionless parameters ($Ka_{\mathrm{f}}^{F}$, $Da_{\mathrm{f}}^{F}$) despite different thickening factors and flow conditions. \\

\midrule

\textit{B}
& $Ka_{\mathrm{f}}\in\{1,4,12\}$,
  $\Lambda_{\mathrm{f}}^{F}=2$,
  $F\in\{1,2,4,8,16,32,64\}$
& $Ka_{\mathrm{f}}$,
  $\Lambda_{\mathrm{f}}^{F}$
& $F$,
  $Ka_{\mathrm{f}}^{F}=\sqrt{F}\,Ka_{\mathrm{f}}$,
  $\Lambda_{\mathrm{f}}=2F$
& Quantify the influence of artificial thickening at fixed prescribed $Ka_{\mathrm{f}}$, while maintaining constant $\Lambda_{\mathrm{f}}^{F}$.\\

\textit{B}$^{\star}$
& $F=1$,
  $Ka_{\mathrm{f}}=Ka_{\mathrm{f}}^{F}$
  of \textit{B},
  $\Lambda_{\mathrm{f}}=2$
& $F=1$,
  $\Lambda_{\mathrm{f}}=2$
& $Ka_{\mathrm{f}}$
& Unthickened reference cases matched to the effective dimensionless parameters ($Ka_{\mathrm{f}}^{F}$, $Da_{\mathrm{f}}^{F}$) of database \textit{B}.\\

\midrule

\textit{L}
& $Ka_{\mathrm{f}}\in\{1,4,12,36\}$,
  $\Lambda_{\mathrm{f}}\in\{2,4,8\}$,
  $F\in\{1,2,4,8,16\}$
& $Ka_{\mathrm{f}}$,
  $\Lambda_{\mathrm{f}}$,
  $\Omega$
& $F$,
  $Ka_{\mathrm{f}}^{F}$,
  $\Lambda_{\mathrm{f}}^{F}=\Lambda_{\mathrm{f}}/F$,
  $L_{\mathrm{domain}}/l_{\mathrm{f}}^{F}
  =\Omega\Lambda_{\mathrm{f}}/F$
& Investigate the effects of decreasing ratios of $L_{\mathrm{int}}$ and $L_{\mathrm{domain}}$ to the thickened flame thickness $l_{\mathrm{f}}^{F}$ with increasing $F$.
Note that $Ka_{\mathrm{f}}^{F}$ and $Da_{\mathrm{f}}^{F}$ also change simultaneously with increasing $F$.\\

\textit{L}$^{\star}$
& $F=1$,
  $Ka_{\mathrm{f}}=Ka_{\mathrm{f}}^{F}$
  of \textit{L},
  $\Lambda_{\mathrm{f}}=\Lambda_{\mathrm{f}}^{F}$
  of \textit{L}
& $F=1$
& $Ka_{\mathrm{f}}$,
  $\Lambda_{\mathrm{f}}$
& Unthickened reference cases matched to the effective dimensionless parameters ($Ka_{\mathrm{f}}^{F}$, $Da_{\mathrm{f}}^{F}$) of database \textit{L}.\\

\textit{L}$^{\mathrm{s}}$
& $F=1$,
  $Ka_{\mathrm{f}}\in\{1,4,12,36\}$,
  $\Lambda_{\mathrm{f}}=1$
& $F=1$,
  $\Lambda_{\mathrm{f}}=1$
& $Ka_{\mathrm{f}}$
& Extend the unthickened reference database ($F=1$ cases included in both \textit{L} and \textit{L}$^\star$) toward smaller $\Lambda_{\mathrm{f}}$.\\

\midrule

\textit{D}
& $F=1$,
  $Ka_{\mathrm{f}}$ as in \textit{L}/\textit{L}$^{\star}$ (reference cases with $F=1$ and $\Lambda_{\mathrm{f}}=2$),
  $\Omega=10$
& $F=1$,
  $Ka_{\mathrm{f}}$,
  $\Omega$
& $Da_{\mathrm{f}}$,
  $\Lambda_{\mathrm{f}}$,
  $L_{\mathrm{int}}$,
  $L_{\mathrm{domain}}$
& Quantify combined effects of varying $L_{\mathrm{int}}$ and $L_{\mathrm{domain}}$ relative to flame thickness $l_{\mathrm{f}}$, while maintaining a constant domain-to-integral-length-scale ratio $\Omega$ at fixed $Ka_{\mathrm{f}}$ (equivalently, $Ka_{\mathrm{f}}^{F}$ for $F=1$).\\

\textit{P}
& $F=1$,
 $Ka_{\mathrm{f}}$ and $Da_{\mathrm{f}}$ as in $\Lambda_{\mathrm{f}}=2$, $F=1$ cases of \textit{L}/\textit{L}$^{\star}$; varying $\Omega$
& $F=1$,
  $Ka_{\mathrm{f}}$,
  $Da_{\mathrm{f}}$,
  $L_{\mathrm{int}}$
& $L_{\mathrm{domain}}$,
  $\Omega$
& Isolate the individual effects of $L_{\mathrm{domain}}$ and $L_{\mathrm{int}}$ at fixed $Ka_{\mathrm{f}}$ by periodically replicating the turbulent forcing field and varying $\Omega$.\\

\bottomrule
\end{tabularx}
\end{table*}

An overview of the databases, including their parameter ranges and respective objectives, is provided in Table~\ref{tab:databases}; a brief description of each database follows below.

Database \textit{A} consists of simulations with different unthickened $Ka_{\mathrm{f}}$, $\Lambda_{\mathrm{f}}$, and thickening factors $F$, chosen such that $\Lambda_{\mathrm{f}}^{F}=2$ and $Ka_{\mathrm{f}}^{F}\in\{1,4,12,36\}$.
According to the scaling relations derived in Sec.~\ref{sec:theory}, cases with identical $Ka_{\mathrm{f}}^{F}$ and $\Lambda_{\mathrm{f}}^{F}$ should exhibit similar flame dynamics after accounting for $F$, despite differences in their prescribed $Ka_{\mathrm{f}}$ and physical domain sizes $L_{\mathrm{domain}}$.

Database \textit{B} comprises several sets of simulations at different fixed values of $Ka_{\mathrm{f}}$. Within each set, the thickening factor $F$ and $\Lambda_{\mathrm{f}}$ (through $L_{\mathrm{domain}}$) are varied such that $\Lambda_{\mathrm{f}}^{F}=2$ is maintained, while $Ka_{\mathrm{f}}^{F}$ varies with $F$.
This database therefore isolates the effect of artificial flame thickening on flame propagation while maintaining a constant ratio between the integral length scale and the thickened flame thickness $\Lambda_{\mathrm{f}}^{F}$.
In addition, database \textit{B}$^{\star}$ comprises unthickened ($F=1$) simulations with $Ka_{\mathrm{f}}=Ka_{\mathrm{f}}^{F}$ and $\Lambda_{\mathrm{f}}=\Lambda_{\mathrm{f}}^{F}$, matching the effective Karlovitz numbers and length-scale ratios of the corresponding thickened-flame cases.

Database \textit{L} is designed to investigate the effect of reducing the turbulent and geometric length scales relative to the thickened flame thickness.
It contains simulations with different $Ka_{\mathrm{f}}$ and $\Lambda_{\mathrm{f}}$, with $F$ varied for each parameter combination.
Because $\Omega=L_{\mathrm{domain}}/L_{\mathrm{int}}$ remains constant, increasing $F$ simultaneously decreases both $L_{\mathrm{domain}}/(Fl_{\mathrm{f}})$ and $\Lambda_{\mathrm{f}}^{F}$.
The most strongly thickened cases reach values as low as $L_{\mathrm{domain}}/(Fl_{\mathrm{f}})=1.25$ and $\Lambda_{\mathrm{f}}^{F}=0.125$, such that $l_{\mathrm{f}}^{F}$ becomes comparable to $L_{\mathrm{domain}}$ and exceeds $L_{\mathrm{int}}$.
These cases extend beyond conditions typically targeted in LES, probe the limiting scale relations discussed in Sec.~\ref{sec:theory}, including regimes relevant to fully modelled turbulence as in RANS, and approach the nominally laminar region of the turbulent combustion regime diagram.
Database \textit{L}$^{\star}$ also contains corresponding $F=1$ simulations with $Ka_{\mathrm{f}}=Ka_{\mathrm{f}}^{F}$ and $\Lambda_{\mathrm{f}}=\Lambda_{\mathrm{f}}^{F}$ chosen to match the effective quantities $Ka_{\mathrm{f}}^{F}$ and $\Lambda_{\mathrm{f}}^{F}$ of the thickened cases.
An additional set, denoted database \textit{L}$^{\mathrm{s}}$, contains only unthickened flames ($F=1$) at Karlovitz numbers corresponding to those considered in databases \textit{L} and \textit{L}$^{\star}$, with $\Lambda_{\mathrm{f}}=\Lambda_{\mathrm{f}}^{F}=1$.
These simulations extend the unthickened reference database toward smaller $L_{\mathrm{int}}$.

Two further unthickened datasets are introduced to distinguish restrictions imposed by the turbulent integral length scale from those associated with the finite domain.
Database \textit{D} contains unthickened simulations at the same $Ka_{\mathrm{f}}^{F}$ values as the $\Lambda_{\mathrm{f}}=2$ cases from database \textit{L} and their corresponding matched cases from database \textit{L}$^{\star}$, but with independently varied $Da_{\mathrm{f}}^{F}$, and hence varied $L_{\mathrm{int}}$ and $L_{\mathrm{domain}}$, while retaining $\Omega = L_{\mathrm{domain}}/L_{\mathrm{int}}\approx 10$.
Database \textit{P}, in contrast, varies $L_{\mathrm{domain}}$ while keeping $L_{\mathrm{int}}$ and thus $Da_{\mathrm{f}}$ fixed.
This is achieved by periodically replicating the turbulent forcing field in the lateral directions.
The resulting simulations therefore permit the effects of geometric domain confinement and integral-length-scale limitations to be separated, as suggested in~\cite{hunt2025}.

\begin{figure}
    \centering
    \includegraphics[width=\linewidth]{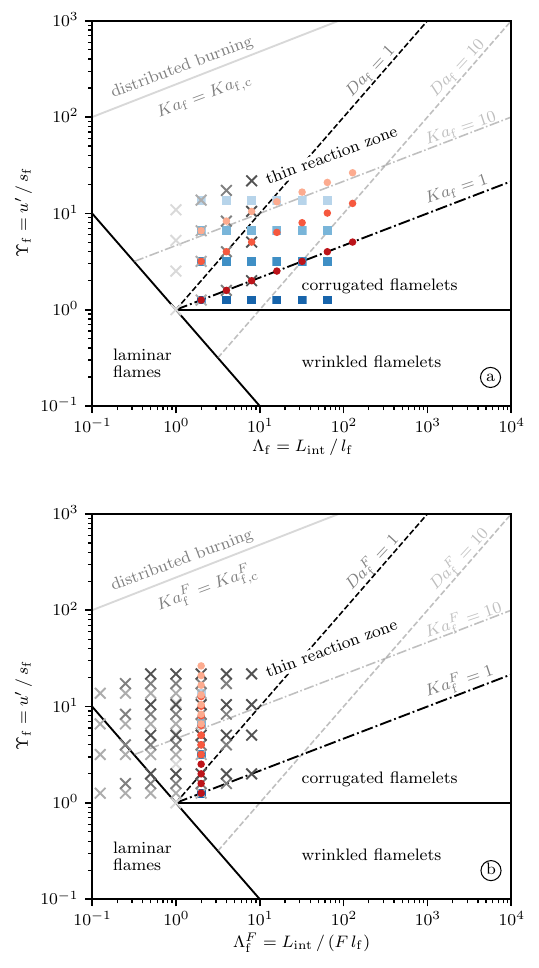}
                        \caption{
        Premixed turbulent combustion regime diagram showing the simulation conditions of databases \textit{A}, \textit{B}, \textit{L}, and \textit{L}$^{\mathrm{s}}$.
        Database \textit{A} (\textcolor{dbBlue}{\raisebox{0.0ex}{\rule{0.5em}{0.5em}}}) is shown in blue, with lighter shades indicating increasing $Ka_{\mathrm{f}}^{F}$; database \textit{B} (\textcolor{dbRed}{\raisebox{0.0ex}{\ensuremath{\bullet}}}) is shown in red, with lighter shades indicating increasing $Ka_{\mathrm{f}}$; and databases \textit{L} and \textit{L}$^{\mathrm{s}}$ (\textcolor{dbGray}{\raisebox{0.0ex}{\normalsize$\times$}}) are shown in grey, with lighter shades indicating decreasing $\Lambda_{\mathrm{f}}$.
    }
    \label{fig:regime_diagram_base}
\end{figure}

Figure~\ref{fig:regime_diagram_base} illustrates how artificial flame thickening shifts the simulation conditions in the turbulent combustion regime diagram through the modification of $\Lambda_{\mathrm{f}}^{F}$, while $\Upsilon_{\mathrm{f}}^{F}=\Upsilon_{\mathrm{f}}$.

The operating and turbulent conditions for the more than 200 cases are provided in Tables~S.1--S.8 of the Supplementary Material.

\section{Results}
\label{sec:results}

The results are first discussed in terms of the effects of artificial flame thickening on local flame propagation, followed by its influence on flame wrinkling.
This separation reflects that the global flame consumption speed can be expressed as the product of contributions from local flame propagation and flame surface area (see Eq.~\eqref{eq:sc_I0_Psi})~\cite{bray1991}.

As discussed in~\cite{meziatRamirez2025,schneider2026c}, an ATF model for \ce{H2} combustion must therefore account for these two coupled contributions separately to accurately predict the overall flame consumption speed.
This is also consistent with the discussion of Lapenna et al.~\cite{lapenna2025}, who emphasised that turbulent combustion models for \ce{H2} must capture unresolved flame wrinkling, differential-diffusion and TD effects, as well as their interaction with turbulence.
Within these two aspects of local flame propagation and flame wrinkling, the effects of the different parametric variations introduced above are discussed successively.

\subsection{Effects of flame thickening on local flame speed}
\label{subsec:local_flame_speed}

The first objective of this section is to assess whether the previously discussed increase in the effective Karlovitz number $Ka_{\mathrm{f}}^{F}$ indeed provides an appropriate description of the local turbulence--flame interaction in artificially thickened flames.
To this end, Sec.~\ref{subsubsec:assessment_effective_Karlovitz} compares the cases from database \textit{A} at fixed $Ka_{\mathrm{f}}^{F}$ but varying thickening factor $F$ in terms of their instantaneous flame-surface structure, mean local flame speed, and distributions of local flame speed conditioned on curvature.
If $Ka_{\mathrm{f}}^{F}$ captures the relevant effect of thickening, these quantities should exhibit comparable behaviour across the corresponding cases.
From a modelling perspective, it is furthermore important to quantify how strongly the mean local flame speed increases with $F$ when the prescribed, unthickened Karlovitz number $Ka_{\mathrm{f}}$ is held fixed.
Thus, Sec.~\ref{subsubsec:scaling_mean_local_flame_speed} examines the cases from databases \textit{B} and \textit{B}$^{\star}$ and compares the resulting trends with literature models developed for unthickened flames to assess whether the thickening-induced increase in local flame speed can be predicted by accounting for the corresponding change in Karlovitz number.
Finally, Sec.~\ref{subsubsec:lengthscale_effects_mean_local_flame_speed} investigates the influence of the turbulence length scales using cases from databases \textit{L}, \textit{L}$^{\star}$, \textit{L}$^{\mathrm{s}}$, \textit{D} and \textit{P}.
First, $F$ is varied at fixed domain size $L_{\mathrm{domain}}$ and therefore also fixed integral length scale $L_{\mathrm{int}}$ (databases \textit{L} and \textit{L}$^{\star}$).
To separate the effects of $L_{\mathrm{domain}}$ and $L_{\mathrm{int}}$ more systematically, additional cases with varying $L_{\mathrm{domain}}$ are considered.
These comprise cases with a fixed scale-separation ratio $\Omega=L_{\mathrm{domain}}/L_{\mathrm{int}}$, for which $L_{\mathrm{int}}$ varies together with the domain size (database \textit{D}), as well as cases with fixed $L_{\mathrm{int}}$ and consequently varying $\Omega$ (database \textit{P}).
This enables the effects of domain size and integral length scale to be assessed as independently as possible.

\subsubsection{Assessment of the effective Karlovitz number}
\label{subsubsec:assessment_effective_Karlovitz}

The effect of artificial flame thickening on local flame propagation is first assessed using database \textit{A}.
The cases were constructed such that, for varying $F=1,2,4,8,16,32$, the effective conditions $Ka_{\mathrm{f}}^{F}=Ka_{\mathrm{f}}\sqrt{F}$ and $\Lambda_{\mathrm{f}}^{F}=\Lambda_{\mathrm{f}}/F=2$, and therefore also $Da_{\mathrm{f}}^{F}$, remain fixed.

\begin{figure*}
    \centering
    \includegraphics[width=\linewidth]{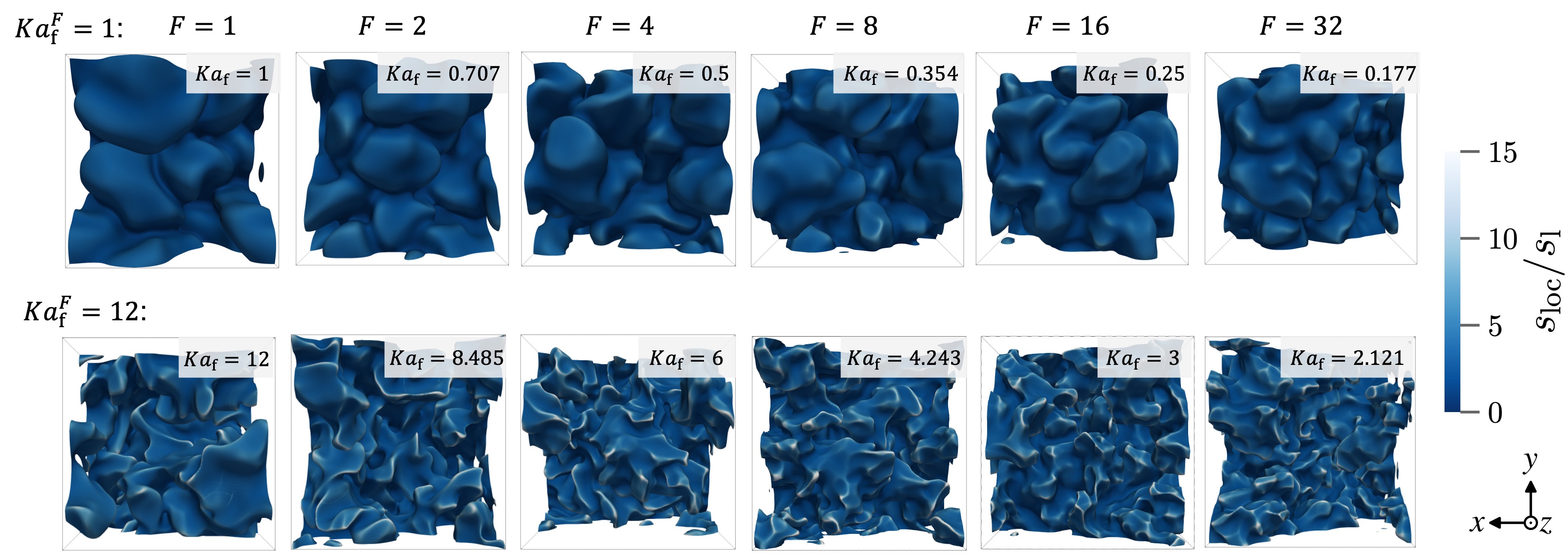}
    \caption{
        Flame surfaces defined by the progress variable ($PV$) at $PV=0.9$ for $Ka_{\mathrm{f}}^{F}=1$ (top) and $Ka_{\mathrm{f}}^{F}=12$ (bottom), coloured by the normalised local flame speed $s_{\mathrm{loc}}/s_{\mathrm{l}}$.
        The thickening factor $F$ increases from left to right, with a corresponding decrease in the prescribed $Ka_{\mathrm{f}}$.
        Spatial coordinates are normalised by the thickened flame thickness, $Fl_{\mathrm{f}}$.
    }
    \label{fig:flame_surfaces}
\end{figure*}

Figure~\ref{fig:flame_surfaces} shows representative flame surfaces for two of the effective Karlovitz numbers considered, $Ka_{\mathrm{f}}^{F}=1$ and $12$.
Corresponding visualisations for all investigated values of $Ka_{\mathrm{f}}^{F}$, along with an additional version using unscaled spatial coordinates for $Ka_{\mathrm{f}}^{F}=12$, are provided in the Supplementary Material (Figs.~S.2 and~S.3).

The local flame speed $s_{\mathrm{loc}}$, used to colour the flame surfaces, (and the local flame thickness $l_{\mathrm{loc}}$) are evaluated using the methodology introduced by Day et al.~\cite{day2009} and subsequently employed in~\cite{howarth2022,howarth2023,hunt2025}.
The $PV=0.9$ isosurface is triangulated, and flame-normal trajectories are traced from the triangle vertices along the local progress-variable gradient to construct the flame-normal tubes, with $PV=1-Y_{\ce{H2}}/Y_{\ce{H2},\mathrm{u}}$.
For the present operating condition, $PV=0.9$ approximately coincides with the location of maximum heat release in the corresponding one-dimensional unstretched freely propagating flame.

Despite the substantial variation in the prescribed Karlovitz number $Ka_{\mathrm{f}}$ within each set of constant $Ka_{\mathrm{f}}^{F}$, the corresponding flame surfaces exhibit a very similar morphology.
For example, for $Ka_{\mathrm{f}}^{F}=4$, $Ka_{\mathrm{f}}$ varies from $0.707$ to $4$, while for $Ka_{\mathrm{f}}^{F}=36$ it varies from $6.364$ to $36$.
The distributions of $s_{\mathrm{loc}}/s_{\mathrm{l}}$ likewise remain very similar at constant $Ka_{\mathrm{f}}^{F}$, while larger values of $Ka_{\mathrm{f}}^{F}$ lead to systematically higher local flame speeds.
Conversely, cases with similar prescribed $Ka_{\mathrm{f}}$ do not exhibit similar flame structures or local propagation characteristics when their thickening factors differ substantially.

Another important implication of artificial flame thickening becomes apparent when considering flame-front structures in unscaled spatial coordinates.
Although the flame-front morphologies appear similar when normalised by the thickened flame thickness (Fig.~\ref{fig:flame_surfaces}), their characteristic length scales increase with $F$ (see also Fig.~S.3 in the Supplementary Material).
For instance, at $F=32$, structures of a given normalised size correspond to physical length scales 32 times larger than in the unthickened case, consistent with observations for laminar flames~\cite{schuh2024}.

\begin{figure}
    \centering
    \includegraphics[width=\linewidth]{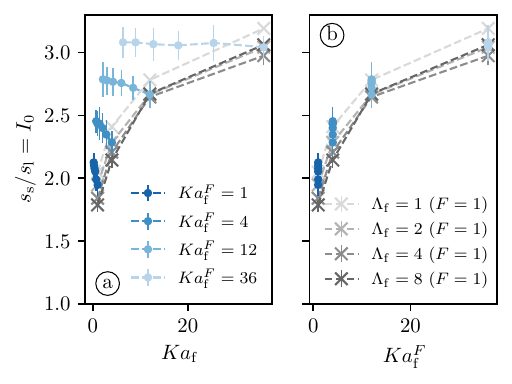}
    \caption{
        (a) normalised mean local flame speed $s_{\mathrm{s}}/s_{\mathrm{l}}=I_0$ as a function of the prescribed Karlovitz number $Ka_{\mathrm{f}}$ for fixed $Ka_{\mathrm{f}}^{F}=1,4,12,36$ and varying thickening factor $F$.
        Dashed grey lines show the $F=1$ cases from datasets \textit{L} and \textit{L}$^{\mathrm{s}}$ for $Ka_{\mathrm{f}}=Ka_{\mathrm{f}}^{F}=1,4,12,36$ and $\Lambda_{\mathrm{f}}=1,2,4,8$.
        (b) Same data plotted as a function of the effective Karlovitz number accounting for flame thickening, $Ka_{\mathrm{f}}^{F}$.
        Vertical bars indicate one standard deviation.
    }
    \label{fig:I0_Ka_f_Ka_fF_Ka_fF_fixed}
\end{figure}

To quantify the effect of flame thickening on local flame propagation, Fig.~\ref{fig:I0_Ka_f_Ka_fF_Ka_fF_fixed}(a) shows the normalised mean local flame speed $s_{\mathrm{s}}/s_{\mathrm{l}}=\overline{s_{{\mathrm{loc}}}}/s_{\mathrm{l}}=I_0$ as a function of the prescribed $Ka_{\mathrm{f}}$ for all cases in dataset \textit{A}.
Figure~\ref{fig:I0_Ka_f_Ka_fF_Ka_fF_fixed}(b) shows the same data as a function of $Ka_{\mathrm{f}}^{F}$.
For reference, the grey lines show the $F=1$ cases from dataset \textit{L} (and \textit{L}$^{\mathrm{s}}$), covering $Ka_{\mathrm{f}}=Ka_{\mathrm{f}}^{F}=1,4,12,36$ and $\Lambda_{\mathrm{f}}=1,2,4,8$.

The $F=1$ reference cases (grey lines of varying brightness) show only a weak dependence of $I_0$ on domain size at fixed $Ka_{\mathrm{f}}$, consistent with~\cite{hunt2025}, while $I_0$ increases with $Ka_{\mathrm{f}}$, as reported in~\cite{howarth2023}.
When plotted against the prescribed $Ka_{\mathrm{f}}$, the deviation of the thickened flames from the $F=1$ reference increases with $F$ (corresponding to a decreasing $Ka_{\mathrm{f}}$ at fixed $Ka_{\mathrm{f}}^{F}$; see Eq.~\eqref{eq:KaF_DaF}).
In contrast, $I_0$ remains approximately constant for each fixed $Ka_{\mathrm{f}}^{F}$, as becomes particularly apparent in Fig.~\ref{fig:I0_Ka_f_Ka_fF_Ka_fF_fixed}(b).
The remaining scatter is comparable to the small scatter observed for the $F=1$ reference flames across variations in $\Lambda_{\mathrm{f}}$.
Thus, to leading order, artificial thickening modifies local flame propagation through the increase in $Ka_{\mathrm{f}}^{F}\propto \sqrt{F}$.

These results also highlight the advantage of characterising turbulence--flame interaction using the flame-scale Karlovitz number $Ka_{\mathrm{f}}$ rather than the conventional Kolmogorov-based definition $Ka_{\eta,\mathrm{f}}=\tau_{\mathrm{f}}/\tau_\eta$, with $\tau_\eta=\sqrt{\nu/\varepsilon}$.
The two Karlovitz-number definitions are related through the flame Reynolds number as $Ka_{\eta,\mathrm{f}}=\sqrt{Re_{\mathrm{f}}}\,Ka_{\mathrm{f}}$, with $Re_{\mathrm{f}}=s_{\mathrm{f}}l_{\mathrm{f}}/\nu$.
Artificial flame thickening increases the flame Reynolds number according to $Re_{\mathrm{f}}^{F}=FRe_{\mathrm{f}}$, because the flame thickness is increased by $F$, while the kinematic viscosity $\nu$ is not modified by thickening.
Consequently, $Ka_{\eta,\mathrm{f}}^{F}=FKa_{\eta,\mathrm{f}}$, in contrast to $Ka_{\mathrm{f}}^{F}=\sqrt{F}\,Ka_{\mathrm{f}}$.
Thus, although the two definitions are related for a given flame, the proportionality between them changes with $F$ because thickening alters $Re_{\mathrm{f}}^{F}$.

This difference is illustrated in Fig.~S.7 of the Supplementary Material.
For cases with fixed $Ka_{\mathrm{f}}^{F}$, $Ka_{\eta,\mathrm{f}}^{F}$ varies substantially with $F$, while $s_{\mathrm{s}}/s_{\mathrm{l}}$ remains approximately constant.
The thickened-flame data therefore do not follow the $Ka_{\eta,\mathrm{f}}$ dependence of the $F=1$ reference flames, supporting $Ka_{\mathrm{f}}$ as the more suitable measure of small-scale turbulence--flame interaction.

Beyond the influence of flame thickening on surface-averaged quantities, its effect on the local flame structure can be assessed from the dependence of the local flame speed on flame curvature.
The flame curvature is defined as $\kappa = \frac{\partial n_j}{\partial x_j}$, where the flame-surface normal vector $n_j$ points toward the unburned gases.

The evaluated curvature depends to some extent on the selected flame isosurface.
For consistency, curvature is evaluated at $PV=0.9$ for all cases.
Since the present analysis primarily concerns relative changes with thickening factor and Karlovitz number rather than absolute curvature values, using the same isosurface provides a suitable basis for comparison.

\begin{figure*}
    \centering
    \includegraphics[width=\linewidth]{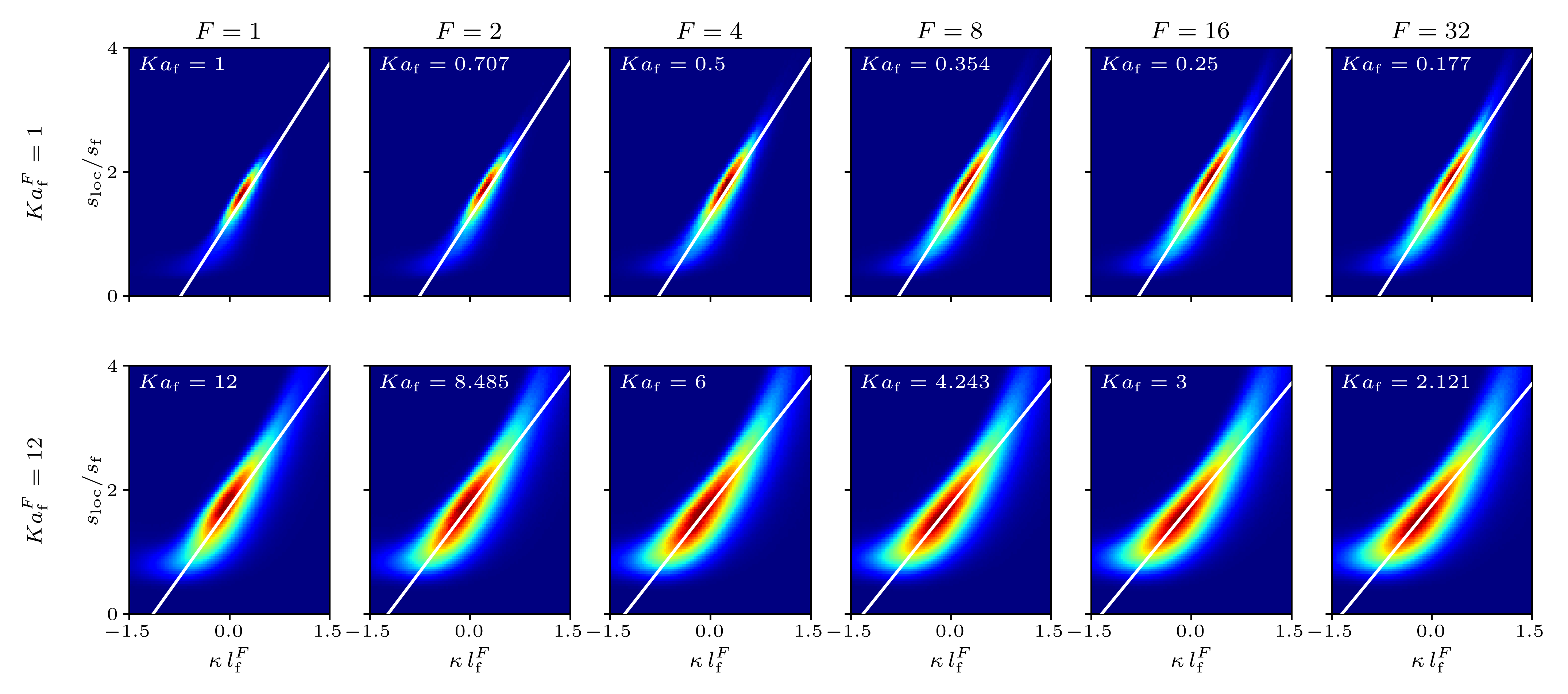}
    \caption{
    Local consumption-speed-weighted joint probability density functions (JPDFs) of the normalised local flame speed $s_{\mathrm{loc}}/s_{\mathrm{f}}$ and normalised curvature $\kappa l_{\mathrm{f}}^{F}$ for $Ka_{\mathrm{f}}^{F}=1$ and $12$ from database \textit{A}.
    White lines indicate linear fits obtained by orthogonal distance regression.
    }
    \label{fig:curv_local_flame_speed_JPDFs}
\end{figure*}

Figure~\ref{fig:curv_local_flame_speed_JPDFs} shows the corresponding distributions of the normalised local flame speed $s_{\mathrm{loc}}/s_{\mathrm{f}}$ as a function of the normalised curvature $\kappa l_{\mathrm{f}}^{F}$ for the cases at $Ka_{\mathrm{f}}^{F}=1$ and $12$ from database \textit{A}, where $l_{\mathrm{f}}^{F}$ denotes the thickened flame thickness.
For each case, the first moment with respect to $s_{\mathrm{loc}}$ is considered, such that the distribution accounts for where the flame is actively burning rather than for the amount of flame surface alone \cite{howarth2022}.
Accordingly, the integral over the distribution is related to the mean local flame speed, rather than representing a measure of flame surface area.
This is particularly relevant because the flame-surface distribution can be strongly influenced by extended regions of weak burning or local extinction, for example at the back of the flame brush; see Howarth et al.~\cite{howarth2023} for further discussion.

Although the cases at a given $Ka_{\mathrm{f}}^{F}$ span a broad range of prescribed, unthickened Karlovitz numbers $Ka_{\mathrm{f}}$, a high degree of similarity is obtained once curvature is normalised by the thickened flame thickness $l_{\mathrm{f}}^{F}$.
In particular, the overall shape and orientation of the distributions remain approximately unchanged as $F$ is increased at constant $Ka_{\mathrm{f}}^{F}$.
Some redistribution of probability within this common structure is nevertheless visible: with increasing $F$, the distributions become slightly less concentrated around their highest-probability region and more broadly distributed along the same overall pattern.
In contrast, clear differences are observed between the two effective Karlovitz numbers.

The orthogonal-distance-regression fits through the local flame-speed distributions show that the response to curvature becomes slightly weaker with increasing $Ka_{\mathrm{f}}^{F}$, as indicated by the decreasing magnitude of the fitted slopes.
This behaviour is consistent with the decreasing sensitivity of the local flame speed to curvature at increasing turbulence intensity reported by Howarth et al.~\cite{howarth2023}.
At fixed $Ka_{\mathrm{f}}^{F}$, the fitted slopes, and hence the corresponding effective consumption-based Markstein numbers characterising the curvature response, remain approximately independent of $F$; for a detailed discussion of consumption-based Markstein numbers and their implications for modelling local flame consumption speeds in unthickened, TD unstable flames, the reader is referred to Howarth et al.~\cite{howarth2022,howarth2023}.
Thus, artificial flame thickening does not introduce a fundamentally different curvature response when curvature is consistently normalised by the thickened flame thickness.
These observations are also consistent with the scaling arguments for artificially thickened flames discussed in~\cite{popp2019,detomaso2023}, according to which the effect of flame stretch $K$, comprising both curvature and tangential strain contributions, on the structure of an artificially thickened flame effectively scales as $F K$.
This follows from the corresponding increase in the characteristic flame thickness and flame time, $l_{\mathrm{f}}^{F}=F l_{\mathrm{f}}$ and $\tau_{\mathrm{f}}^{F}=F\tau_{\mathrm{f}}$, respectively.

The previously identified scaling with $Ka_{\mathrm{f}}^{F}$ therefore extends beyond surface-averaged quantities to the local flame structure.
Overall, increasing the thickening factor $F$ while simultaneously decreasing the prescribed Karlovitz number $Ka_{\mathrm{f}}$ such that the effective Karlovitz number $Ka_{\mathrm{f}}^{F}$ remains constant produces a similar local flame response.

\subsubsection{Scaling of mean local flame speed with flame thickening}
\label{subsubsec:scaling_mean_local_flame_speed}

Next, simulation database \textit{B} is used to quantify the increase in mean local flame speed with $F$ at fixed prescribed $Ka_{\mathrm{f}}$.

Here, $Ka_{\mathrm{f}}\in\{1,4,12\}$ is fixed while $F$ is varied from 1 to 64.
To also avoid restricting TD effects through the available length scale, the ratio of the integral length scale to the thickened flame thickness is kept constant at $\Lambda_{\mathrm{f}}^{F}=2$.

Using Eq.~\eqref{eq:ss_sf_fit_Kaf} and the relation $Ka_{\mathrm{f}}^{F}=Ka_{\mathrm{f}}\sqrt{F}$ from Eq.~\eqref{eq:KaF_DaF} yields
\begin{equation}
    \begin{aligned}
        \frac{s_{\mathrm{s}}}{s_{\mathrm{f}}}
        &= 1 + 0.26 e^{-0.038\omega_2}
        \sqrt{Ka_{\mathrm{f}}^{F}} \\
        &= 1 + 0.26 e^{-0.038\omega_2}
        \sqrt{Ka_{\mathrm{f}}}\,F^{1/4}\,.
    \end{aligned}
    \label{eq:ss_sf_Ka_scaling_with_F}
\end{equation}
Hence, at fixed $Ka_{\mathrm{f}}$, the increase in mean local flame speed is expected to scale as $s_{\mathrm{s}}/s_{\mathrm{f}}-1\sim F^{1/4}$.

\begin{figure}
    \centering
    \includegraphics[width=\linewidth]{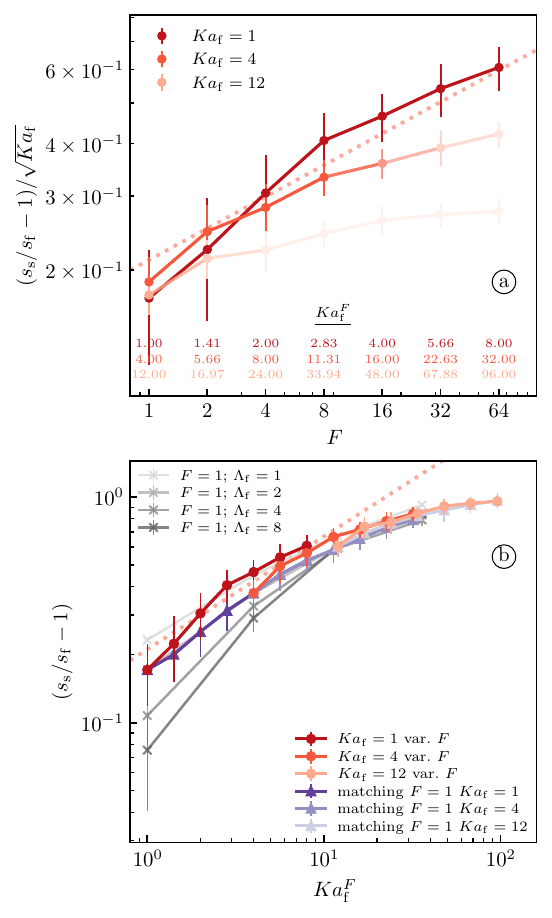}
    \caption{
        (a) $(s_{\mathrm{s}}/s_{\mathrm{f}}-1)/\sqrt{Ka_{\mathrm{f}}}$ as a function of $F$ for fixed $Ka_{\mathrm{f}}=1,4,12$.
        The dotted line shows Eq.~\eqref{eq:ss_sf_Ka_scaling_with_F}, rearranged accordingly, corresponding to the $F^{1/4}$ scaling.
        Lighter colours at fixed $Ka_{\mathrm{f}}$ indicate cases with $Ka_{\mathrm{f}}^{F}>12$.
        (b) $s_{\mathrm{s}}/s_{\mathrm{f}}-1$ as a function of $Ka_{\mathrm{f}}^{F}$ for the same cases. 
        Corresponding $F=1$ reference cases with matching $Ka_{\mathrm{f}}^{F}$ and $Da_{\mathrm{f}}^{F}$ (dataset \textit{B}$^{\star}$), as well as $F=1$ cases from datasets \textit{L} and \textit{L}$^{\mathrm{s}}$ ($\Lambda_{\mathrm{f}}=1,2,4,8$), are included for comparison.
        The dotted line shows Eq.~\eqref{eq:ss_sf_Ka_scaling_with_F}, rearranged accordingly.
        Vertical bars indicate one standard deviation; although generally small, they become more apparent in the logarithmic representation.
    }
    \label{fig:Ka_f_fixed_over_F_Ka_fF}
\end{figure}

Figure~\ref{fig:Ka_f_fixed_over_F_Ka_fF}(a) shows $(s_{\mathrm{s}}/s_{\mathrm{f}}-1)/\sqrt{Ka_{\mathrm{f}}}$ as a function of $F$ on logarithmic axes for the three prescribed Karlovitz numbers.
With this normalisation, Eq.~\eqref{eq:ss_sf_Ka_scaling_with_F} reduces to a single $F^{1/4}$ dependence for all $Ka_{\mathrm{f}}$.
The corresponding values of $Ka_{\mathrm{f}}^{F}$ are indicated below the individual data points.

For $Ka_{\mathrm{f}}^{F}\lesssim12$, the simulations follow the predicted $F^{1/4}$ scaling closely and largely independently of the prescribed $Ka_{\mathrm{f}}$.
At larger $Ka_{\mathrm{f}}^{F}$, however, the increase in $s_{\mathrm{s}}$ with $F$ becomes weaker.
Figure~\ref{fig:Ka_f_fixed_over_F_Ka_fF}(b) shows that this behaviour results from the underlying dependence of the local flame speed on $Ka_{\mathrm{f}}^{F}$: above approximately $Ka_{\mathrm{f}}^{F}=12$, the simulations increasingly deviate from the correlation of Howarth et al.~\cite{howarth2023}.
This trend is observed not only for the thickened-flame cases from dataset \textit{B}, but also for the corresponding $F=1$ reference cases, including dataset \textit{B}$^{\star}$ and the simulations at $F=1$ with varying $\Lambda_{\mathrm{f}}$ from datasets \textit{L} and \textit{L}$^{\mathrm{s}}$. It is also consistent with the results of Howarth et al.~\cite{howarth2023} for the same operating condition considered here.

Thus, the $F^{1/4}$ scaling provides a good description as long as the corresponding $Ka_{\mathrm{f}}^{F}$ remains within the range in which $s_{\mathrm{s}}/s_{\mathrm{f}}-1\propto\sqrt{Ka_{\mathrm{f}}^{F}}$.
At larger $Ka_{\mathrm{f}}^{F}$, the increase in mean local flame speed with $F$ is weaker than $F^{1/4}$.

Additionally, it should be noted that $s_{\mathrm{l}}$ is substantially smaller than $s_{\mathrm{f}}$.
Consequently, relative deviations in the predicted turbulence-induced enhancement, $s_{\mathrm{s}}/s_{\mathrm{f}}-1$, are more pronounced than the corresponding deviations relative to the total $I_0=s_{\mathrm{s}}/s_{\mathrm{l}}$.
Thus, although the present representation emphasises discrepancies at higher $Ka_{\mathrm{f}}^{F}$, the model still provides reasonably accurate predictions of $I_0$ over a somewhat wider range of Karlovitz numbers.

\subsubsection{Length-scale effects on mean local flame speed}
\label{subsubsec:lengthscale_effects_mean_local_flame_speed}

The following analysis uses database \textit{L} to examine the regime in which the thickened flame thickness $l_{\mathrm{f}}^{F}$ becomes comparable to the characteristic turbulent and geometric length scales.
With increasing $F$, $l_{\mathrm{f}}^{F}$ first approaches and may exceed $L_{\mathrm{int}}$ and, for stronger thickening, also the lateral domain size $L_{\mathrm{domain}}$, which bounds the maximum attainable integral length scale.
As discussed in Sec.~\ref{sec:theory}, such scale relations are particularly relevant to the coarse resolutions encountered in RANS and URANS.
Database $\textit{L}^{\star}$ provides corresponding unthickened ($F=1$) reference cases matched in the effective parameters, including $Ka_{\mathrm{f}}^{F}$ and $\Lambda_{\mathrm{f}}^{F}$.

\begin{figure}
    \centering
    \includegraphics[width=\linewidth]{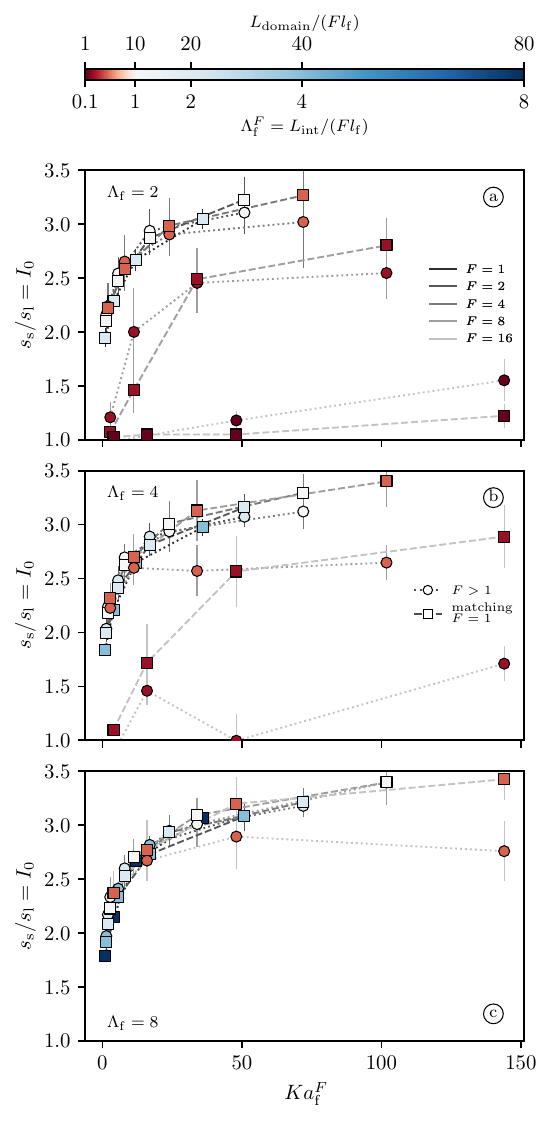}
    \caption{
        Normalised mean local flame speed $s_{\mathrm{s}}/s_{\mathrm{l}}=I_0$ as a function of $Ka_{\mathrm{f}}^{F}$ for database \textit{L} at (a) $\Lambda_{\mathrm{f}}=2$, (b) $\Lambda_{\mathrm{f}}=4$, and (c) $\Lambda_{\mathrm{f}}=8$.
        Colours indicate the effective length-scale ratio $\Lambda_{\mathrm{f}}^{F}=L_{\mathrm{int}}/(F l_{\mathrm{f}})$, which is directly related to $L_{\mathrm{domain}}/(F l_{\mathrm{f}})$ through $\Omega=L_{\mathrm{domain}}/L_{\mathrm{int}}=10$.
        Dots denote cases from database \textit{L}, while squares denote the corresponding $F=1$ reference cases from database \textit{L}$^{\star}$.
        Vertical bars indicate one standard deviation.
    }
    \label{fig:I0_Ka_fF_Lambda_f_conditioned_greys_topcolorbar}
\end{figure}

Figure~\ref{fig:I0_Ka_fF_Lambda_f_conditioned_greys_topcolorbar} shows that, as long as $\Lambda_{\mathrm{f}}^{F}\gtrsim1$, the simulations collapse onto a common dependence of $s_{\mathrm{s}}/s_{\mathrm{l}} = I_0$ on $Ka_{\mathrm{f}}^{F}$ across different $F$ and $\Lambda_{\mathrm{f}}$.
Once $\Lambda_{\mathrm{f}}^{F}$ decreases below approximately unity, $I_0$ decreases at fixed $Ka_{\mathrm{f}}^{F}$ and approaches $I_0\approx1$ (corresponding to $s_{\mathrm{s}}\approx s_{\mathrm{l}}$) for sufficiently small $\Lambda_{\mathrm{f}}^{F}$.
Note that, because $\Omega=L_{\mathrm{domain}}/L_{\mathrm{int}}=10$ is a property of the forcing, $\Lambda_{\mathrm{f}}^{F}=1$ corresponds to $L_{\mathrm{domain}}/(F l_{\mathrm{f}})=10$.
In terms of the turbulent combustion regime diagram, increasing $F$ at fixed $\Lambda_{\mathrm{f}}$ can therefore be interpreted as a continuous shift toward the laminar-flame regime as the scale separation between the flame thickness and the integral length scale is lost (see Fig.~\ref{fig:regime_diagram_base}).

For cases with $\Lambda_{\mathrm{f}}^{F}<1$, increasing $Ka_{\mathrm{f}}^{F}$ at fixed $\Lambda_{\mathrm{f}}$ and $F$ generally increases $s_{\mathrm{s}}/s_{\mathrm{l}}$.
In the turbulent combustion regime diagram, this corresponds to moving upward from the laminar-flame region toward more turbulence-dominated regimes.
Although $L_{\mathrm{int}}$ remains unchanged, increasing $Ka_{\mathrm{f}}^{F}$ through larger turbulent velocity fluctuations ($u^{\prime}$) increases the turbulent kinetic energy, including that associated with scales comparable to the thickened flame thickness ($l_{\mathrm{f}}^{F}$), even when these scales exceed $L_{\mathrm{int}}$.
Consequently, turbulent fluctuations at scales comparable to the thickened flame thickness, even when exceeding $L_{\mathrm{int}}$, can contribute more strongly to the flame dynamics.
This behaviour is illustrated schematically in Fig.~S.9 of the Supplementary Material. Figure~S.11(b) also shows that $I_0$ increases modestly with increasing $Ka_{\mathrm{f}}^{F}$ (and hence $u^{\prime}$), including when $L_{\mathrm{int}}/(F l_{\mathrm{f}})<1$.
Note that this behaviour is not restricted to artificially thickened flames: although flame thickening facilitates reaching conditions with $\Lambda_{\mathrm{f}}^{F}<1$, similar conditions may also occur in flames without artificial thickening, for example when the nozzle diameter of a jet flame is small relative to the flame thickness.

For $\Lambda_{\mathrm{f}}^{F}<1$, the thickened-flame cases and their nominally matched $F=1$ reference cases show similar qualitative trends but do not necessarily yield identical values.
Once $\Lambda_{\mathrm{f}}^{F}\lesssim1$, $Ka_{\mathrm{f}}^{F}$ remains algebraically well defined but loses its usual interpretation as a measure of turbulence at the flame scale, since $l_{\mathrm{f}}^{F}$ no longer lies within the inertial subrange and instead exceeds the integral length scale.
Consequently, matching $Ka_{\mathrm{f}}^{F}$ and $Da_{\mathrm{f}}^{F}$ between thickened and $F=1$ flames does not necessarily imply equivalent turbulence--flame interaction in this regime.

Additional JPDFs of the local flame speed and curvature are provided in Fig.~S.8 of the Supplementary Material.
As $\Lambda_{\mathrm{f}}^{F}$ decreases, the correlations between these quantities progressively weaken, the corresponding distributions flatten, and the characteristic increase in local flame speed becomes strongly reduced.

A property of the forcing employed in database \textit{L}, and thus a limitation for the present analysis, is the approximately fixed ratio between domain size and integral length scale, $\Omega=L_{\mathrm{domain}}/L_{\mathrm{int}}\approx10$, as described above.
Consequently, these simulations alone do not allow the individual effects of $L_{\mathrm{domain}}$ and $L_{\mathrm{int}}$ on the local flame-speed enhancement to be separated.
However, based on previous studies \cite{aspden2017a,schuh2024,altantzis2012,berger2019,frouzakis2015}, both length scales ($L_{\mathrm{int}}$ and $L_{\mathrm{domain}}$) are expected to play a role.

If $L_{\mathrm{int}}$ becomes comparable to or smaller than the thickened flame thickness, the dominant energy-containing turbulent scales are smaller than the flame thickness, thereby reducing their influence on local flame structure and propagation~\cite{aspden2017a}.
Independently, studies of laminar freely propagating flames have shown that sufficiently small domains restrict both global and local flame-speed enhancement, with the flame speed approaching its one-dimensional laminar value as the domain size is reduced~\cite{altantzis2012,frouzakis2015,berger2019,schuh2024}.
This reduction typically becomes relevant when $L_{\mathrm{domain}}$ approaches the critical wavelength $\lambda_{\mathrm{crit}}$, obtained from a linear stability analysis, which marks the transition between stable and unstable perturbation wavelengths in the dispersion relation~\cite{altantzis2012}.

To separate these effects as far as possible, two additional simulation databases are considered.
In database \textit{D}, the domain size is increased relative to the $\Lambda_{\mathrm{f}}=2$, $F=1$ baseline cases from database \textit{L}$^{\star}$, while retaining $\Omega\approx10$.
Hence, $L_{\mathrm{domain}}$ and $L_{\mathrm{int}}$ increase simultaneously, corresponding to increasing $Da_{\mathrm{f}}$ at fixed $Ka_{\mathrm{f}}$.
In the turbulent combustion regime diagram, this corresponds to moving diagonally upward and to the right along a line of constant $Ka_{\mathrm{f}}$.

Database \textit{P} employs periodic forcing, as described in Sec.~\ref{sec:num_setup} and in~\cite{hunt2025}, to vary $L_{\mathrm{domain}}$ independently of $L_{\mathrm{int}}$.
Thus, $L_{\mathrm{int}}$ remains fixed while $L_{\mathrm{domain}}$, and consequently $\Omega$, is varied.
The cases therefore correspond to the same location in the turbulent combustion regime diagram while independently varying the geometric confinement.
The relationship between $L_{\mathrm{domain}}$ and $L_{\mathrm{int}}$ for database \textit{P} is illustrated in Fig.~S.10 of the Supplementary Material, together with the corresponding cases from database \textit{D} and those from database \textit{L}$^{\star}$ matched to the $\Lambda_{\mathrm{f}}=2$ cases from database \textit{L} for reference.
Note that, for each case in database \textit{D}, a matching case exists in database \textit{P} with identical $L_{\mathrm{domain}}$ and prescribed $Ka_{\mathrm{f}}$, but a different $L_{\mathrm{int}}$ and, consequently, $Da_{\mathrm{f}}$.

\begin{figure}
    \centering
    \includegraphics[width=\linewidth]{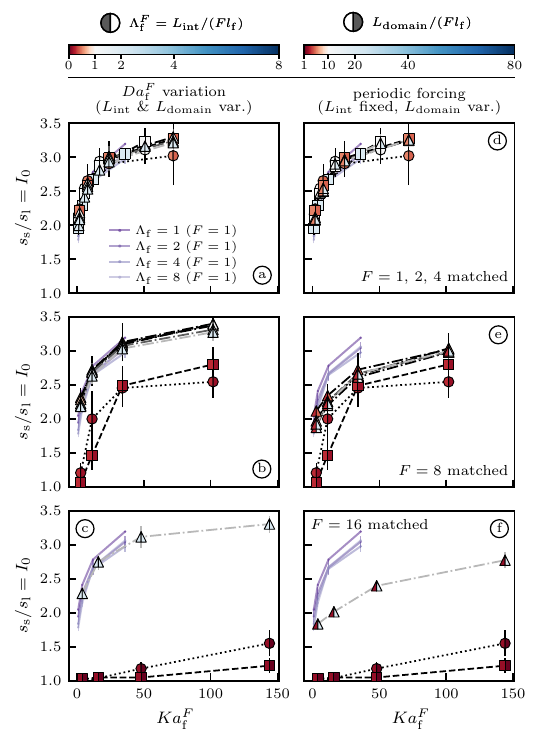}
    \caption{
        Normalised mean local flame speed, $s_{\mathrm{s}}/s_{\mathrm{l}}=I_0$, as a function of $Ka_{\mathrm{f}}^{F}$ for variations in $L_{\mathrm{domain}}$ with either varying or fixed $L_{\mathrm{int}}$.
        Left column (a-c): database \textit{D}, for which $L_{\mathrm{domain}}$ and $L_{\mathrm{int}}$ are increased simultaneously at constant $\Omega$.
        Right column (d-f): database \textit{P}, for which $L_{\mathrm{domain}}$ is varied at fixed $L_{\mathrm{int}}$.
        Rows correspond to $F=1,2,4$ (top), $F=8$ (middle), and $F=16$ (bottom).
        Cases from database \textit{L} at $\Lambda_{\mathrm{f}}=2$ and the corresponding $F=1$ baseline cases from database \textit{L}$^{\star}$ are included for reference.
        Marker halves indicate $\Lambda_{\mathrm{f}}^{F}$ (left) and $L_{\mathrm{domain}}/(F l_{\mathrm{f}})$ (right), respectively.
        Additionally, $F=1$ cases from databases \textit{L} and \textit{L}$^{\mathrm{s}}$ ($\Lambda_{\mathrm{f}}^{F}=\Lambda_{\mathrm{f}}=1,2,4,8$) are included as purple lines in all panels for reference.
        Vertical error bars indicate one standard deviation.
    }
    \label{fig:I0_Ka_fF_split_marker_LambdaF_LdomainF_greys}
\end{figure}

Figure~\ref{fig:I0_Ka_fF_split_marker_LambdaF_LdomainF_greys} compares the effects of simultaneously increasing $L_{\mathrm{domain}}$ and $L_{\mathrm{int}}$ (varying $Da_{\mathrm{f}}$; left column, a–c) with those of increasing only $L_{\mathrm{domain}}$ at fixed $L_{\mathrm{int}}$ (constant $Da_{\mathrm{f}}$; right column, d–f).
The matched $F=1$ cases from database \textit{L}$^{\star}$, designed to correspond to the $\Lambda_{\mathrm{f}}=2$ cases from database \textit{L}, serve as the respective baselines.
For the $F=1$, $2$, and $4$ cases and the corresponding matched $F=1$ cases, length-scale effects are generally small because the flame thickness remains sufficiently separated from the integral and geometric scales.
Only at the highest Karlovitz numbers is a weak increase in $I_0$ observed with increasing $L_{\mathrm{domain}}$ and, respectively, $L_{\mathrm{int}}$.
Consequently, the differences between databases \textit{D} and \textit{P} remain small for these cases, consistent with the limited influence of geometric confinement and the integral length scale for $L_{\mathrm{domain}}/l_{\mathrm{f}}^{F}\geq5$ and $\Lambda_{\mathrm{f}}^{F}\geq0.5$.

For $F=8$, the influence of the available length scales becomes considerably stronger, as already evident from Fig.~\ref{fig:I0_Ka_fF_Lambda_f_conditioned_greys_topcolorbar}(b).
As $Da_{\mathrm{f}}^{F}$ increases, the associated increases in both $L_{\mathrm{domain}}$ and $L_{\mathrm{int}}$ rapidly restore $I_0$ to values approaching those obtained without significant length-scale restrictions, as also indicated by comparison with the purple $F=1$ reference lines for $\Lambda_{\mathrm{f}}=1,2,4,8$.
In contrast, increasing $L_{\mathrm{domain}}$ alone while keeping $L_{\mathrm{int}}$ fixed produces only a moderate increase in $I_0$ and does not recover the values obtained for $\Lambda_{\mathrm{f}}^{F}\gtrsim1$, but instead tends toward its own asymptotic limit (compare Fig.~\ref{fig:I0_Ka_fF_Lambda_f_conditioned_greys_topcolorbar}(b,e)).
This behaviour becomes even more pronounced for $F=16$ (Fig.~\ref{fig:I0_Ka_fF_Lambda_f_conditioned_greys_topcolorbar}(c,f)).

These results indicate that both $L_{\mathrm{domain}}$ and $L_{\mathrm{int}}$ can limit and modify the mean local flame-speed enhancement, albeit through different mechanisms.

In general, when $L_{\mathrm{domain}}$ is limiting, increasing it progressively removes geometric confinement and thereby increases $I_0$, consistent with observations for intrinsically unstable laminar flames~\cite{schuh2024} and with the corresponding three-dimensional laminar results presented in the Supplementary Material (Fig.~S.12).
When $L_{\mathrm{int}}$ is fixed significantly below the thickened flame thickness ($\Lambda_{\mathrm{f}}^{F}\ll1$) and $L_{\mathrm{domain}}$ is also sufficiently small to limit the flame, increasing $L_{\mathrm{domain}}$ progressively removes this geometric restriction.
The resulting flame then approaches the behaviour of a freely propagating TD-unstable laminar flame, exhibiting a characteristic intrinsically unstable flame-front structure and an associated increase in $I_0$ to values approaching $s_{\mathrm{f}}$.
In this regime, turbulent fluctuations at scales comparable to or larger than the flame thickness are weak because $L_{\mathrm{int}}<F l_{\mathrm{f}}$, such that the flame dynamics are predominantly governed by the intrinsic TD instability rather than by flame-scale turbulence, consistent with the spectra shown in Fig.~\ref{fig:energy_spectra}.
Hence, an integral length scale $L_{\mathrm{int}}$ smaller than the thickened flame thickness introduces an additional length-scale restriction that cannot be eliminated by increasing the domain size $L_{\mathrm{domain}}$ alone.
Increasing the turbulent velocity fluctuations ($u^{\prime}$) at fixed $L_{\mathrm{int}}$ enhances turbulent motions at scales relevant to the flame (see schematic Fig.~S.9 in the Supplementary Material), resulting in a modest increase in $I_0$ relative to the nearly laminar, intrinsically unstable state (see Fig.~S.11(b) in the Supplementary Material).

When $Da_{\mathrm{f}}$ is increased by simultaneously increasing $L_{\mathrm{int}}$ and $L_{\mathrm{domain}}$, both restrictions are progressively removed.
First, the increase in $L_{\mathrm{domain}}$ alleviates geometric confinement, analogous to the behaviour observed for the periodically forced cases.
At the same time, increasing $L_{\mathrm{int}}$ shifts the flame thickness back into the inertial subrange.
Consequently, the assumptions underlying the definition of $Ka_{\mathrm{f}}^{F}$ become applicable again, and turbulent fluctuations can interact with the flame at scales comparable to its thickness in a manner consistent with the underlying inertial-range scaling.
The corresponding increase in $I_0$ therefore reflects the recovery of turbulence--flame interactions at the flame scale in addition to the removal of geometric confinement.

Since $\Omega=L_{\mathrm{domain}}/L_{\mathrm{int}}\geq10$, the characteristic domain length scale $L_{\mathrm{domain}}$ is always substantially larger than the integral length scale $L_{\mathrm{int}}$.
Consequently, the present database does not include cases in which the influence of $L_{\mathrm{domain}}$ can be isolated while the flame thickness remains within the inertial subrange and flame-scale turbulence--chemistry interaction is retained.

\begin{figure}
    \centering
    \includegraphics[width=\linewidth]{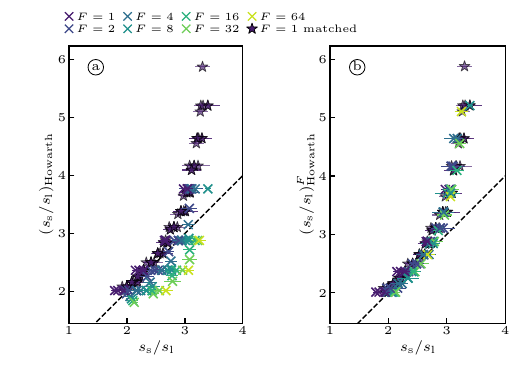}
    \caption{
    Model prediction of the normalised mean local flame speed $(s_{\mathrm{s}}/s_{\mathrm{l}})_{\mathrm{Howarth}}$ versus the normalised mean local flame speed $s_{\mathrm{s}}/s_{\mathrm{l}}$ obtained from the simulations:
    (a) $(s_{\mathrm{s}}/s_{\mathrm{l}})_{\mathrm{Howarth}}$ evaluated using the prescribed Karlovitz number $Ka_{\mathrm{f}}$ (Eq.~\eqref{eq:ss_sf_fit_Kaf}), and
    (b) $(s_{\mathrm{s}}/s_{\mathrm{l}})_{\mathrm{Howarth}}^{F}$ evaluated using the effective Karlovitz number $Ka_{\mathrm{f}}^{F}$ (Eq.~\eqref{eq:ss_sf_Ka_scaling_with_F}).
    Cases with $\Lambda_{\mathrm{f}}^{F}<1$ are excluded.
    Horizontal lines indicate the standard deviation obtained from the simulations.
    }
    \label{fig:ss_sl_Howarth_parity_unthickened_vs_thickened}
\end{figure}

Overall, the results discussed above show that, as long as the thickened flame scale remains smaller than the integral length scale ($\Lambda_{\mathrm{f}}^{F}\geq1$), accounting for flame thickening in the Karlovitz and Damk\"ohler numbers through $Ka_{\mathrm{f}}^{F}$ and $Da_{\mathrm{f}}^{F}$ leads to a good collapse of the normalised mean local flame speed across the different cases.
Furthermore, the mean local flame speed is well described by the model of Howarth et al.~\cite{howarth2023} up to approximately $Ka_{\mathrm{f}}^{F}=12$.

This behaviour is summarised in Fig.~\ref{fig:ss_sl_Howarth_parity_unthickened_vs_thickened} using all cases with $\Lambda_{\mathrm{f}}^{F}\geq1$.
When the prescribed $Ka_{\mathrm{f}}$ is used in Fig.~\ref{fig:ss_sl_Howarth_parity_unthickened_vs_thickened}(a), the cases increasingly deviate from the overall scatter with increasing $F$.
In contrast, using the effective $Ka_{\mathrm{f}}^{F}$ in Fig.~\ref{fig:ss_sl_Howarth_parity_unthickened_vs_thickened}(b) largely removes this systematic dependence on $F$.
For $s_{\mathrm{s}}/s_{\mathrm{l}}\gtrsim3$, corresponding approximately to $Ka_{\mathrm{f}}^{F}\gtrsim12$, the model increasingly overpredicts the mean local flame speed; this deviation reflects a limitation of the model rather than an effect of flame thickening.

\subsection{Effects of flame thickening on flame wrinkling}
\label{subsec:ft_effects_wrinkling}

The effect of artificial flame thickening on flame wrinkling $\Psi$ is considered next.
For TD unstable flames, the analysis examines whether flame wrinkling follows a consistent scaling when both TD effects on the local flame speed and the thickening-induced changes in the governing nondimensional parameters are accounted for, and whether established wrinkling models remain applicable under these conditions.
As discussed in Sec.~\ref{sec:theory}, the Damköhler number is a key parameter governing flame wrinkling, including the transition between the small-scale and large-scale limits~\cite{damkoehler1940,hunt2025}.

\subsubsection{Assessment of thickening-adjusted scaling for flame wrinkling}

For the SSL, Eq.~\eqref{eq:ssl_lsl_peters} can equivalently be expressed in terms of $\Lambda_r$ and $Ka_r$ as:
\begin{equation}
    \Psi
    =
    1
    +
    \gamma_r Ka_r^{1/3}\Lambda_r^{2/3},
    \label{eq:psi_Ka_Lambda}
\end{equation}
using the coefficients $\gamma_r$ reported in~\cite{hunt2025,hunt2025_phd}.
These coefficients differ from those originally proposed by Peters~\cite{peters1999}.
As discussed by Hunt and Aspden~\cite{hunt2025}, the discrepancy appears to arise primarily from differences in the definitions of the underlying dimensionless quantities (that is, how TD effects are accounted for in those definitions) rather than from low-Lewis-number effects alone, although a residual dependence on the fuel Lewis number remains.

\begin{figure}
    \centering
    \includegraphics[width=\linewidth]{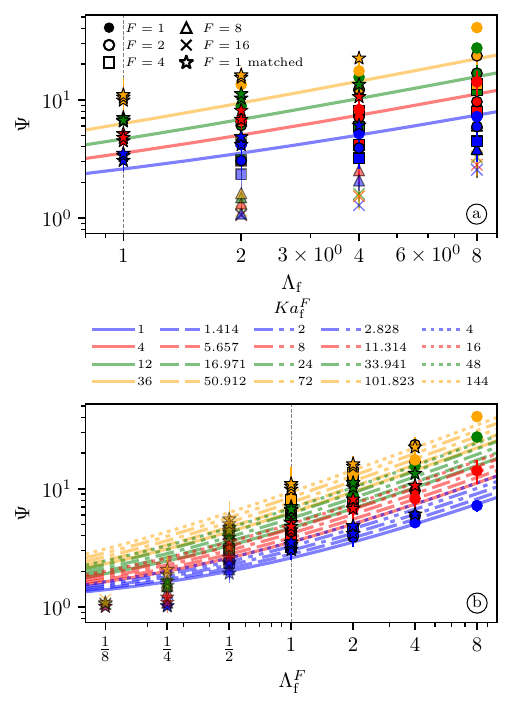}
    \caption{
        Flame wrinkling $\Psi$ as a function of (a) $\Lambda_{\mathrm{f}}$ and (b) $\Lambda_{\mathrm{f}}^{F}$ for simulations from databases \textit{L} and \textit{L}$^{\star}$.
        Symbols distinguish the thickening factor $F$, while stars denote the matched $F=1$ cases from database \textit{L}$^{\star}$.
        Colours indicate the corresponding $Ka_{\mathrm{f}}$ in (a) and $Ka_{\mathrm{f}}^{F}$ in (b).
        Cases with $\Lambda_{\mathrm{f}}^{F}<1$ (equivalently $L_{\mathrm{domain}}/(F l_{\mathrm{f}})<10$) are shown with reduced opacity.
        Lines with different linestyles denote the SSL model in Eq.~\eqref{eq:psi_Ka_Lambda} for the Karlovitz numbers indicated in the legend.
        $\Lambda_{\mathrm{f}}=1$ (a) and $\Lambda_{\mathrm{f}}^{F}=1$ (b) are indicated by grey vertical dashed lines.
        Vertical bars indicate one standard deviation.
    }
    \label{fig:Psi_Lambda_f_Lambda_fF}
\end{figure}

Figure~\ref{fig:Psi_Lambda_f_Lambda_fF} shows $\Psi$ for simulations from databases \textit{L} and \textit{L}$^{\star}$ as a function of $\Lambda_{\mathrm{f}}$ in panel (a) and $\Lambda_{\mathrm{f}}^{F}$ in panel (b).
Accordingly, the SSL curves in panel (a) are evaluated for the different $Ka_{\mathrm{f}}$ values, whereas those in panel (b) use the corresponding $Ka_{\mathrm{f}}^{F}$ values.

When $\Psi$ is plotted against the prescribed length-scale ratio $\Lambda_{\mathrm{f}}$, the deviation of the thickened-flame cases from the SSL model generally increases with $F$.
Accounting for flame thickening through $\Lambda_{\mathrm{f}}^{F}$ and $Ka_{\mathrm{f}}^{F}$ substantially improves the collapse, as shown in Fig.~\ref{fig:Psi_Lambda_f_Lambda_fF}(b), for example for the cases at fixed $Ka_{\mathrm{f}}=1$.

The scaling remains reasonably representative down to $\Lambda_{\mathrm{f}}^{F}\approx1$ and, for most cases, even to $\Lambda_{\mathrm{f}}^{F}\approx0.5$.
At smaller values, however, the wrinkling progressively approaches $\Psi=1$, although a weak increase with $Ka_{\mathrm{f}}^{F}$ remains.
This behaviour is consistent with the loss of scale separation discussed above in the context of local flame propagation (Sec.~\ref{subsubsec:lengthscale_effects_mean_local_flame_speed}): as the thickened flame thickness exceeds the integral length scale, the range of turbulent scales available to wrinkle the flame becomes increasingly restricted.

Corresponding results for databases \textit{A} and \textit{B} are provided in the Supplementary Material (Fig.~S.13) and show the same overall trends.
Across the databases, a small systematic deviation from the SSL model remains, suggesting that a slightly larger value of $\gamma_{\mathrm{f}}$ would improve quantitative agreement.
This is discussed further below.

\begin{figure}
    \centering
    \includegraphics[width=\linewidth]{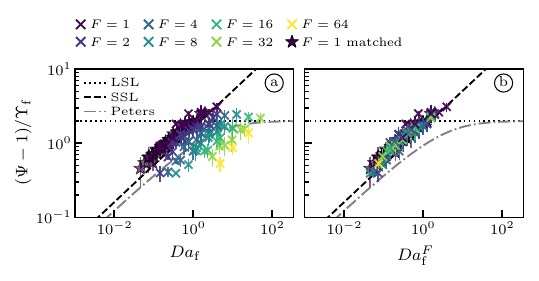}
    \caption{
        Normalised flame-wrinkling enhancement $(\Psi-1)/\Upsilon_{\mathrm{f}}$ as a function of (a) $Da_{\mathrm{f}}$ and (b) $Da_{\mathrm{f}}^{F}$, coloured by the thickening factor $F$.
        Simulations corresponding to matched unthickened reference cases ($F=1$), i.e., cases with the same $Ka_{\mathrm{f}}^{F}$ and $Da_{\mathrm{f}}^{F}$ as a thickened flame case, are indicated by $\star$ markers.
        The SSL, LSL, and Peters-model predictions from Eq.~\eqref{eq:ssl_lsl_peters} are additionally shown using the model constants, including $\gamma_{\mathrm{f}}$, reported in~\cite{hunt2025_phd}.
    }
    \label{fig:Psi_SSL_f_normalization}
\end{figure}

The comparison with the established wrinkling models in Eq.~\eqref{eq:ssl_lsl_peters} is subsequently extended using additional data from the other simulation databases.
The normalised enhancement in flame wrinkling is analysed as a function of the Damköhler number, with particular emphasis on whether the thickening-induced change in $Da_{\mathrm{f}}$ consistently accounts for the observed variation in wrinkling.

Figure~\ref{fig:Psi_SSL_f_normalization} shows the normalised flame-wrinkling enhancement $(\Psi-1)/\Upsilon_{\mathrm{f}}$ as a function of $Da_{\mathrm{f}}$ in panel~(a) and $Da_{\mathrm{f}}^{F}$ in panel~(b) for all datasets except \textit{D} and \textit{P}, which are presented in Fig.~S.14 in the Supplementary Material.
Cases with $\Lambda_{\mathrm{f}}^{F}<1$ are excluded from the analysis because, as discussed above, they deviate in both mean local flame speed and flame wrinkling once the thickened flame scale exceeds the integral scale, and are of limited relevance for practical LES applications.

When plotted against the prescribed $Da_{\mathrm{f}}$ in Fig.~\ref{fig:Psi_SSL_f_normalization}(a), the deviations from the corresponding $F=1$ cases increase systematically, and approximately linearly, with increasing $F$.
For the unthickened flames ($F=1$), the data already follow Damköhler's SSL scaling well, consistent with Hunt and Aspden~\cite{hunt2025} and the trends observed in Fig.~\ref{fig:Psi_Lambda_f_Lambda_fF}.
Accounting for flame thickening through $Da_{\mathrm{f}}^{F}$ substantially reduces the scatter and collapses the data closely onto the SSL scaling in Fig.~\ref{fig:Psi_SSL_f_normalization}(b).
This indicates that, also for the TD-unstable hydrogen flames considered here, the leading-order effect of artificial flame thickening on flame wrinkling is captured by the reduction in the effective Damköhler number $Da_{\mathrm{f}}^{F}$.
The Peters flame-wrinkling model~\cite{peters1999}, which blends between the SSL and LSL, also provides a reasonable description of the data.
However, the present simulations remain predominantly within the SSL regime even where the model already predicts a transition toward the LSL, suggesting that the blending may be too strong or occur too early for the conditions considered here.
This highlights the remaining uncertainty regarding the Damköhler number at which this transition occurs, which may also depend on the reactant conditions.

Note that the SSL scaling describes the data well even for the largest configuration considered, with $L_{\mathrm{domain}}=80\,l_{\mathrm{f}}$ and $Da_{\mathrm{f}}^{F}=4$.
For all thickened cases ($F>1$), $Da_{\mathrm{f}}^{F}$ is lower owing to its inverse scaling with $F$.
Larger scale separations and higher Damköhler numbers would therefore be required to investigate the transition toward the LSL, which is beyond the scope of the present study.

The quantities in Fig.~\ref{fig:Psi_SSL_f_normalization} are normalised using the freely propagating flame thickness $l_{\mathrm{f}}$; the corresponding normalisation based on the laminar flame thickness $l_{\mathrm{l}}$ is provided in Fig.~S.16 in the Supplementary Material and yields the same qualitative trends, analogous to~\cite{hunt2025}.

\begin{figure}
    \centering
    \includegraphics[width=\linewidth]{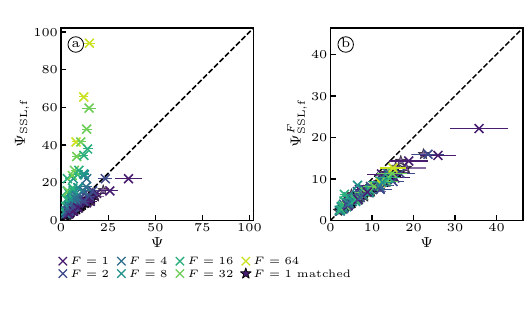}
    \caption{
        Comparison of the flame wrinkling $\Psi$ obtained from the simulations with the SSL-model prediction (Eq.~\eqref{eq:ssl_lsl_peters}) based on (a) the prescribed $Ka_{\mathrm{f}}$ and $Da_{\mathrm{f}}$, $\Psi_{\mathrm{SSL,f}}$, and (b) the thickening-adjusted $Ka_{\mathrm{f}}^{F}$ and $Da_{\mathrm{f}}^{F}$, $\Psi_{\mathrm{SSL,f}}^{F}$.
        Results are shown for databases \textit{A}, \textit{B}, and \textit{L}, including the matched unthickened ($F=1$) cases \textit{B}$^{\star}$ and \textit{L}$^{\star}$ with the same $Ka_{\mathrm{f}}^{F}$ and $Da_{\mathrm{f}}^{F}$ as the corresponding thickened cases.
        Cases with $\Lambda_{\mathrm{f}}^{F}<1$ are excluded.
        Horizontal error bars indicate the standard deviation of $\Psi$.
        The SSL model is evaluated using the model constant $\gamma_{\mathrm{f}}=1.60$ reported in~\cite{hunt2025_phd}.
        The dashed lines indicate $\Upsilon=\Upsilon_{\mathrm{SSL,f}}$ in (a) and $\Upsilon=\Upsilon_{\mathrm{SSL,f}}^{F}$ in (b) for reference.
    }
    \label{fig:Psi_SSL_parity_f_based_reference}
\end{figure}

Figure~\ref{fig:Psi_SSL_parity_f_based_reference} compares the SSL-predicted flame wrinkling $\Psi_{\mathrm{SSL,f}}$ with the corresponding simulation results $\Psi$.
Panel~(a) shows $\Psi_{\mathrm{SSL,f}}$ evaluated using the prescribed $Ka_{\mathrm{f}}$ and $Da_{\mathrm{f}}$, whereas panel~(b) shows $\Psi_{\mathrm{SSL,f}}^{F}$ using the thickening-adjusted effective $Ka_{\mathrm{f}}^{F}$ and $Da_{\mathrm{f}}^{F}$.
Using the prescribed quantities leads to substantial deviations, with the SSL model increasingly overpredicting the actual flame wrinkling as $F$ increases.
Accounting for flame thickening substantially reduces the scatter, although the SSL model slightly underpredicts $\Psi$, as indicated by the systematic deviation from the identity line.

The results in Fig.~\ref{fig:Psi_SSL_parity_f_based_reference} likewise use $\gamma_{\mathrm{f}}=1.60$ from~\cite{hunt2025_phd}.
A least-squares fit to the present database yields $\gamma_{\mathrm{f},\mathrm{opt.}}=2.202$ and substantially improves agreement with the identity line; the corresponding plot (Fig.~S.18) is provided in the Supplementary Material.
Following \cite{hunt2025}, corresponding results based on $l_{\mathrm{l}}$-normalised quantities and an optimised $\gamma_{\mathrm{l,opt.}}$ are also provided in the Supplementary Material (Figs.~S.17 and~S.19) and exhibit very similar trends.

\subsubsection{Scaling of flame wrinkling with flame thickening}

Motivated by the collapse of the wrinkling data obtained when consistently accounting for the thickening factor $F$ in the relevant normalised quantities, as well as by the agreement with the SSL wrinkling model, the effect of increasing $F$ on flame wrinkling $\Psi$ at fixed prescribed $Ka_{\mathrm{f}}$ and $Da_{\mathrm{f}}$ is analysed in more detail, with particular focus on its scaling with $F$.

\begin{figure}
    \centering
    \includegraphics[width=\linewidth]{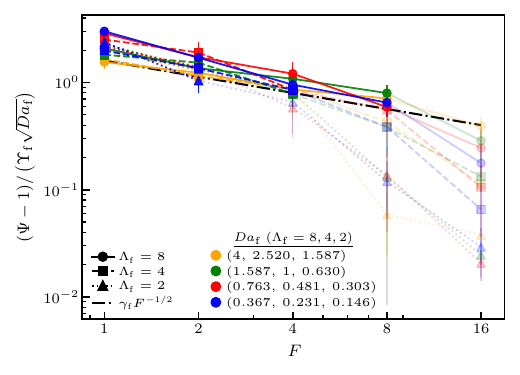}
    \caption{
    Normalised flame-wrinkling enhancement $(\Psi-1)/(\Upsilon_{\mathrm{f}}\sqrt{Da_{\mathrm{f}}})$ as a function of the thickening factor $F$ for all cases from database \textit{L}.
    Cases with $\Lambda_{\mathrm{f}}^{F}<1$, corresponding to $L_{\mathrm{domain}}/(F l_{\mathrm{f}})<10$, are shown for reference with reduced opacity.
    Vertical error bars indicate the standard deviation obtained from the simulations.
    The black dash-dotted line shows the SSL scaling with $\gamma_{\mathrm{f}}=1.60$ from~\cite{hunt2025_phd}.
    }
    \label{fig:Psi_f_Lambda2_4_8_over_F_scaled_with_sqrt_Daf}
\end{figure}

Using $Da_{\mathrm{f}}^{F}=Da_{\mathrm{f}}/F$ (Eq.~\eqref{eq:KaF_DaF}) in the SSL relation of Eq.~\eqref{eq:ssl_lsl_peters} yields
\begin{equation}
    \frac{\Psi-1}{\Upsilon_{\mathrm{f}}\sqrt{Da_{\mathrm{f}}}}
    =
    \gamma_{\mathrm{f}}F^{-1/2} \, .
    \label{eq:Psi_SSL_with_F}
\end{equation}
This relation is examined for database \textit{L} in Fig.~\ref{fig:Psi_f_Lambda2_4_8_over_F_scaled_with_sqrt_Daf}.
The cases with $\Lambda_{\mathrm{f}}^{F}\geq1$ closely follow the predicted $F^{-1/2}$ scaling using $\gamma_{\mathrm{f}}=1.60$; the fitted value $\gamma_{\mathrm{f}}=2.202$ obtained above would shift the prediction slightly upward.
This agreement further supports the scaling $Da_{\mathrm{f}}^{F}=Da_{\mathrm{f}}/F$ and the applicability of the SSL model to the TD-unstable thickened flames considered here.
Large deviations become pronounced once $\Lambda_{\mathrm{f}}^{F}<1$, where the thickened flame thickness $l_{\mathrm{f}}^{F}$ exceeds the integral length scale $L_{\mathrm{int}}$.

These results further suggest that, in the LSL, where $(\Psi-1)/\Upsilon_{\mathrm{f}}=\alpha_{\mathrm{f}}$ is independent of the Damköhler number, artificial flame thickening should not modify the flame wrinkling as long as the thickened flame remains within the LSL.
Since $\Upsilon_{\mathrm{f}}$ is unaffected by thickening, this corresponds to an $F^{0}$ scaling of the wrinkling enhancement.
However, sufficiently strong thickening may reduce $Da_{\mathrm{f}}^{F}$ such that the flame transitions from the LSL toward the SSL.
In this transitional regime, the dependence of wrinkling on $F$ is expected to lie between the LSL scaling, $F^{0}$, and the SSL scaling, $F^{-1/2}$.
Assessing this behaviour will require future simulations at substantially larger Damköhler numbers (corresponding to larger scale separations), such that the unthickened reference flames clearly reside within the LSL.

\subsection{Effects of flame thickening on flame consumption speed}
\label{subsec:ft_effects_sc}

The preceding analyses indicate that the influence of artificial flame thickening on the flame dynamics can be described through the corresponding changes in the effective $Ka_{\mathrm{f}}^{F}$ and $Da_{\mathrm{f}}^{F}$.
To assess this directly, simulations from databases \textit{L} and \textit{L}$^{\star}$, as well as \textit{B} and \textit{B}$^{\star}$, are compared at matching effective $Ka_{\mathrm{f}}^{F}$ and $Da_{\mathrm{f}}^{F}$.
Specifically, each artificially thickened flame with $F>1$ is compared with an unthickened ($F=1$) reference flame at the same effective $Ka_{\mathrm{f}}^{F}$ and $Da_{\mathrm{f}}^{F}$.

\begin{figure}
    \centering
    \includegraphics[width=\linewidth]{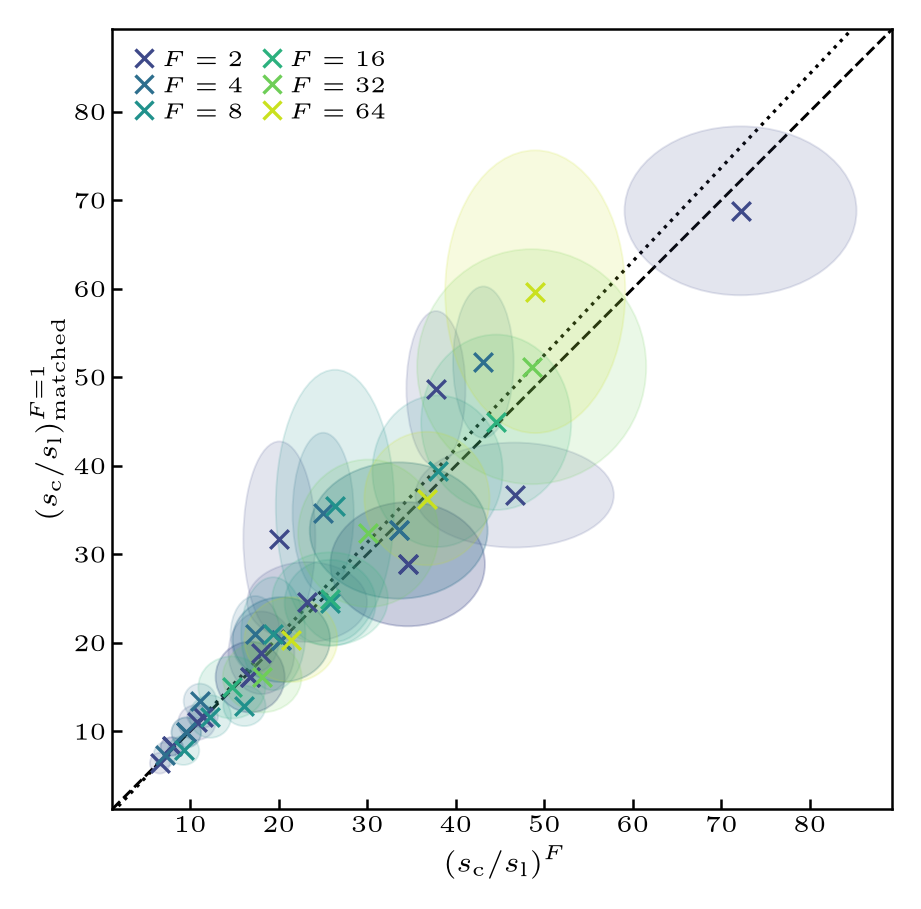}
    \caption{Comparison of the global consumption speed of artificially thickened flames, $(s_{\mathrm{c}}/s_{\mathrm{l}})^{F}$, with that of the corresponding matched unthickened $(F=1)$ simulations, $(s_{\mathrm{c}}/s_{\mathrm{l}})_{\mathrm{matched}}^{F=1}$.
    Colours indicate the thickening factor $F$.
    Symbols denote temporal mean values and ellipses indicate one standard deviation in both directions.
    The dashed line marks exact agreement, and the dotted line shows the linear orthogonal-distance-regression fit.}
    \label{fig:sc_comparison_F_F1}
\end{figure}

Figure~\ref{fig:sc_comparison_F_F1} directly compares the global consumption speeds of the artificially thickened flames with those of the corresponding matched unthickened simulations.
The symbols represent temporal mean values, while the ellipses indicate the temporal variability of both quantities.

The data show very good agreement with the identity line ($(s_{\mathrm{c}}/s_{\mathrm{l}})^{F} = (s_{\mathrm{c}}/s_{\mathrm{l}})^{F=1}_{\mathrm{matched}}$), which is also reflected by the orthogonal-distance-regression fit.
Hence, flames characterised by the same effective $Ka_{\mathrm{f}}^{F}$ and $Da_{\mathrm{f}}^{F}$ exhibit similar propagation behaviour, irrespective of whether these conditions result from artificial flame thickening or from the corresponding unthickened turbulent configuration.
This supports the interpretation of the thickening-induced changes in flame propagation in terms of the modified $Ka_{\mathrm{f}}^{F}$ and $Da_{\mathrm{f}}^{F}$.

As established by the preceding analyses, artificial flame thickening affects these two quantities ($Ka_{\mathrm{f}}$ and $Da_{\mathrm{f}}$) in opposite directions; increasing $F$ increases the effective Karlovitz number while decreasing the effective Damköhler number according to Eq.~\eqref{eq:KaF_DaF}.
Accordingly, flame thickening also affects the two contributions to the global consumption speed in opposite directions.
For the low-to-moderate Karlovitz numbers considered here, the mean local flame speed increases approximately as ${s_{\mathrm{s}}}/{s_{\mathrm{l}}}=I_0\sim\left(Ka_{\mathrm{f}}^{F}\right)^{1/2}\sim F^{1/4}$, whereas in the SSL regime the flame-wrinkling enhancement decreases as $\Psi-1\sim\left(Da_{\mathrm{f}}^{F}\right)^{1/2}\sim F^{-1/2}$.
The net effect of artificial flame thickening on the global consumption speed therefore results from the competition between these two trends.

Since $s_{\mathrm{c}}=\Psi s_{\mathrm{s}}$, the associated changes ($\Psi \downarrow$; $s_{\mathrm{s}} \uparrow$) partially counteract each other but do not generally cancel.
Combining Eq.~\eqref{eq:ss_sf_Ka_scaling_with_F} with the SSL scaling from Eq.~\eqref{eq:Psi_SSL_with_F}, which can be rewritten as $\Psi=1+\gamma_{\mathrm{f}}Ka_{\mathrm{f}}^{F}Da_{\mathrm{f}}^{F}=1+\gamma_{\mathrm{f}}Ka_{\mathrm{f}}Da_{\mathrm{f}}F^{-1/2}$, yields
\begin{align}
\frac{s_{\mathrm{c}}}{s_{\mathrm{l}}}
&=
\Psi
\frac{s_{\mathrm{s}}}{s_{\mathrm{f}}}
\frac{s_{\mathrm{f}}}{s_{\mathrm{l}}}
\nonumber
=
\underbrace{
\frac{s_{\mathrm{f}}}{s_{\mathrm{l}}}
}_{\substack{\text{baseline FP}\\\text{flame speed}}}
+
\underbrace{
\frac{s_{\mathrm{f}}}{s_{\mathrm{l}}}
A\sqrt{Ka_{\mathrm{f}}}F^{1/4}
}_{\substack{\text{local flame-speed}\\\text{enhancement}}} \nonumber
\\
&+
\underbrace{
\frac{s_{\mathrm{f}}}{s_{\mathrm{l}}}
\gamma_{\mathrm{f}}Ka_{\mathrm{f}}Da_{\mathrm{f}}F^{-1/2}
}_{\substack{\text{flame}\\\text{wrinkling}}}
+
\underbrace{
\frac{s_{\mathrm{f}}}{s_{\mathrm{l}}}
A\gamma_{\mathrm{f}}Ka_{\mathrm{f}}^{3/2}Da_{\mathrm{f}}F^{-1/4}
}_{\substack{\text{coupling of local speed}\\\text{and wrinkling}}} \, ,
\label{eq:sc_scaling_F}
\end{align}
with $A=0.26\exp(-0.038\omega_2)$.

The resulting dependence of $s_{\mathrm{c}}$ on $F$ therefore reflects the competition between enhanced local flame speed and reduced flame wrinkling.

\begin{figure}
    \centering
    \includegraphics[width=\linewidth]{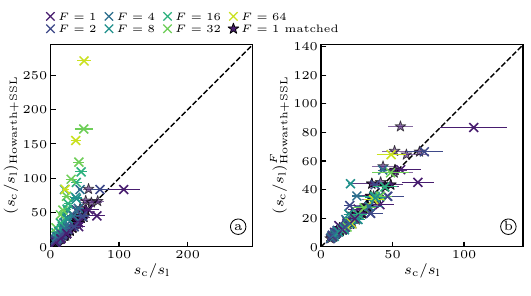}
    \caption{
    Parity plot comparing model predictions of the normalised consumption speed with simulation results. In both panels, the horizontal axis shows the simulation result $s_{\mathrm{c}}/s_{\mathrm{l}}$ and the vertical axis shows the model prediction.
    (a) Predictions $(s_{\mathrm{c}}/s_{\mathrm{l}})_{\mathrm{Howarth+SSL}}$ from Eq.~\eqref{eq:sc_scaling_F} using the prescribed $Ka_{\mathrm{f}}$ and $Da_{\mathrm{f}}$; (b) predictions $(s_{\mathrm{c}}/s_{\mathrm{l}})_{\mathrm{Howarth+SSL}}^{F}$ accounting for flame thickening through the effective $Ka_{\mathrm{f}}^{F}$ and $Da_{\mathrm{f}}^{F}$.
    Horizontal error bars indicate one standard deviation in the simulation results.
    }
    \label{fig:sc_sl_Howarth_SSL_parity_unthickened_vs_thickened}
\end{figure}

Figure~\ref{fig:sc_sl_Howarth_SSL_parity_unthickened_vs_thickened} compares the normalised consumption speed predicted by Eq.~\eqref{eq:sc_scaling_F}, which combines the mean local flame-speed model in Eq.~\eqref{eq:ss_sf_fit_Kaf} with the SSL wrinkling model in Eq.~\eqref{eq:ssl_lsl_peters}, with the corresponding simulation results.
When the prescribed $Ka_{\mathrm{f}}$ and $Da_{\mathrm{f}}$ are used in Fig.~\ref{fig:sc_sl_Howarth_SSL_parity_unthickened_vs_thickened}(a), the cases increasingly deviate from the overall scatter and the identity line with increasing $F$.
Accounting for flame thickening through the effective $Ka_{\mathrm{f}}^{F}$ and $Da_{\mathrm{f}}^{F}$ largely removes this systematic dependence on $F$ and yields good agreement with the simulations (see Fig.~\ref{fig:sc_sl_Howarth_SSL_parity_unthickened_vs_thickened}(b)).
The corresponding comparison using the mean local flame-speed model of Chu et al.~\cite{chu2025} and the wrinkling model of Peters~\cite{peters1999} is shown in the Supplementary Material (Fig.~S.20).
It shows similar trends and good overall agreement with the simulations after accounting for thickening, with a slight tendency to underpredict $s_{\mathrm{c}}/s_{\mathrm{l}}$ consistent with the underprediction of flame wrinkling obtained using the literature value of $\gamma_{\mathrm{f}}$ for the present cases.
This demonstrates that, for the TD-unstable flames considered here, the combined effects of artificial flame thickening on local flame speed and flame wrinkling can be captured by the literature models when the thickening-adjusted effective Karlovitz and Damköhler numbers are used.
Note that the increase in local flame speed with Karlovitz number is expected to cease at sufficiently high $Ka_{\mathrm{f}}$ and may eventually reverse~\cite{yao2024a}, while the present wrinkling scaling is restricted to the SSL regime.
Extending the local flame-speed scaling to higher Karlovitz numbers and accounting for the transition toward the LSL therefore remain topics for future work.

Based on this agreement with the literature models, the effect of flame thickening on the global consumption speed is examined next using the combined model for mean local flame speed and flame wrinkling.
For fixed prescribed turbulence conditions, as relevant to the application of the ATF model in a given geometry and flow configuration, the ratio $s_{\mathrm{c}}(F)/s_{\mathrm{c}}(F=1)$ is obtained by normalising Eq.~\eqref{eq:sc_scaling_F} by its value at $F=1$.
This representation allows the competing contributions of local flame-speed enhancement and reduced flame wrinkling to be assessed as functions of $Ka_{\mathrm{f}}$, $Da_{\mathrm{f}}$, and $F$.

\begin{figure}
    \centering
    \includegraphics[width=\linewidth]{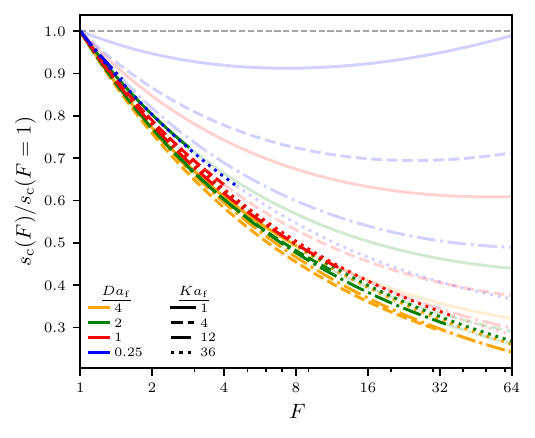}
    \caption{
        Predicted consumption speed normalised by its unthickened value, $s_{\mathrm{c}}(F)/s_{\mathrm{c}}(F=1)$, as a function of the thickening factor $F$ for different $Ka_{\mathrm{f}}$ and $Da_{\mathrm{f}}$.
        Reduced opacity indicates conditions with $\Lambda_{\mathrm{f}}^{F}=Ka_{\mathrm{f}}Da_{\mathrm{f}}^{3/2}/F\leq1$, for which the thickened flame thickness exceeds the turbulent integral length scale.
    }
    \label{fig:sc_scaling_F}
\end{figure}

Figure~\ref{fig:sc_scaling_F} shows the resulting predictions for different $Ka_{\mathrm{f}}$ and $Da_{\mathrm{f}}$, using $\omega_2=5.45$ in the mean local flame-speed model, as determined for the present operating condition.

At sufficiently large $Da_{\mathrm{f}}$, flame wrinkling contributes substantially to the (unthickened) consumption speed, and its reduction with increasing $F$ therefore results in a pronounced decrease in $s_{\mathrm{c}}(F)/s_{\mathrm{c}}(F=1)$.
This effect becomes stronger with increasing $Da_{\mathrm{f}}$ and $Ka_{\mathrm{f}}$ because the SSL wrinkling contribution scales with $Ka_{\mathrm{f}}Da_{\mathrm{f}}$, whereas the local flame-speed enhancement scales only with $\sqrt{Ka_{\mathrm{f}}}$.
At low $Da_{\mathrm{f}}$ and $Ka_{\mathrm{f}}$, the competition between these effects can instead produce a non-monotonic dependence on $F$: initially, the reduction in flame wrinkling dominates, whereas at larger $F$, $\Psi$ approaches unity and the increasing local flame speed can result in a minimum followed by a recovery of $s_{\mathrm{c}}$.

The predicted high-$F$ behaviour at low $Ka_{\mathrm{f}}$ and $Da_{\mathrm{f}}$ must, however, be interpreted with caution, as these conditions correspond to $\Lambda_{\mathrm{f}}^{F}=Ka_{\mathrm{f}}Da_{\mathrm{f}}^{3/2}/F<1$ (these are therefore shown with reduced opacity in Fig.~\ref{fig:sc_scaling_F}).
The minimum and subsequent increase therefore mainly reflect the competition between the two model contributions and are not expected to occur for practically relevant thickening factors.

The same data are shown on logarithmic axes in the Supplementary Material (Fig.~S.21).
The log--log representation shows that all cases with $\Lambda_{\mathrm{f}}^{F}\geq1$ fall approximately between the $F^{-1/2}$ and $F^{-1/4}$, corresponding to the scaling of the wrinkling and coupling terms in Eq.~\eqref{eq:sc_scaling_F}, respectively.

For comparison, the analysis is repeated in the Supplementary Material (Fig.~S.22) using the extended mean local flame-speed model of Chu et al.~\cite{chu2025} and the wrinkling model of Peters~\cite{peters1999}.
These models extend the description toward higher Karlovitz numbers and the transition between the SSL and LSL, respectively; their applicability to artificially thickened TD-unstable hydrogen flames over relevant ranges of $Ka_{\mathrm{f}}$, $Da_{\mathrm{f}}$, and $F$ remains to be assessed in future work.

A direct comparison with corresponding simulation data is also provided in the Supplementary Material (Fig.~S.23).
Overall, the predicted dependence of $s_{\mathrm{c}}(F)/s_{\mathrm{c}}(F=1)$ on the thickening factor is reproduced well by the simulations, although the individual ratios exhibit considerable statistical uncertainty due to temporal fluctuations in flame consumption speed, an inherent feature of turbulent flame propagation.

Note that in this comparison, only thickened simulations for which an unthickened ($F=1$) reference flame is available at the same prescribed $Ka_{\mathrm{f}}$ and $Da_{\mathrm{f}}$ can be considered, i.e., only cases from database \textit{L}.
As discussed above, a broader quantitative assessment would require additional simulations at larger $Da_{\mathrm{f}}$ and $\Lambda_{\mathrm{f}}$; such cases are computationally demanding because increasing the integral length scale requires correspondingly larger computational domains.
A systematic extension of the database toward these conditions is therefore left for future work.

\subsection{Implications for modelling}
\label{subsec:implications_modeling}

The preceding results establish how artificial flame thickening modifies the effective Karlovitz and Damköhler numbers and, consequently, the flame consumption speed through changes in both the mean local flame speed and flame wrinkling.
These findings provide a basis for correcting the thickening-induced modification of flame propagation in coupled simulations.
In particular, such a correction must simultaneously account for the increase in $Ka_{\mathrm{f}}^{F}$ and the decrease in $Da_{\mathrm{f}}^{F}$ and thereby compensate for their combined effect on the flame consumption speed discussed in Sec.~\ref{subsec:ft_effects_sc} and illustrated in Fig.~\ref{fig:sc_scaling_F}.
The objective of the efficiency correction is therefore not to recover the original $Ka_{\mathrm{f}}$ and $Da_{\mathrm{f}}$, but rather to compensate for their combined influence on the resulting flame consumption speed $s_{\mathrm{c}}$.

A direct approach is to introduce an efficiency factor $E$ that compensates for the thickening-induced change in consumption speed, analogous to the formulation proposed by Colin et al.~\cite{colin2000} for conditions under which the effect of thickening is attributed solely to flame wrinkling.
As described in Sec.~\ref{sec:theory}, the efficiency factor increases the effective flame propagation speed according to $s_{\mathrm{f}}^{E}=E s_{\mathrm{f}}$, with the aim of recovering the consumption speed of the unthickened flame.

If feedback effects of the increased flame speed induced by $E$ on the underlying flame--turbulence interaction are neglected, the most straightforward definition is
\begin{equation}
    E
    =
    \frac{s_{\mathrm{c}}(F=1)}
         {s_{\mathrm{c}}(F)},
\end{equation}
where both the numerator and denominator may be evaluated using Eq.~\eqref{eq:sc_scaling_F}.
This effectively reverses the thickening-induced modification of the consumption speed discussed in the context of Fig.~\ref{fig:sc_scaling_F}.

However, the efficiency factor itself modifies the nondimensional parameters governing the turbulence--flame interaction.
Since $E$ increases the effective flame speed, the corresponding thickened quantities become
\begin{equation}
    \Upsilon_{\mathrm{f}}^{F,E}
    =
    \frac{\Upsilon_{\mathrm{f}}}{E} \, ,
    \qquad
    \Lambda_{\mathrm{f}}^{F,E}
    =
    \frac{\Lambda_{\mathrm{f}}}{F} \, .
\end{equation}
Consequently, the effective Karlovitz and Damköhler numbers become
\begin{equation}
    Ka_{\mathrm{f}}^{F,E}
    =
    \frac{\sqrt{F}}{E^{3/2}}Ka_{\mathrm{f}}\, ,
    \qquad
    Da_{\mathrm{f}}^{F,E}
    =
    \frac{E}{F}Da_{\mathrm{f}} \, .
    \label{eq:Ka_Da_FE}
\end{equation}
Thus, application of $E$ introduces an additional shift in the combustion-regime diagram: artificial thickening reduces $\Lambda_{\mathrm{f}} > \Lambda_{\mathrm{f}}^{F} (F>1) = \Lambda_{\mathrm{f}}^{F,E} (F>1; E>1)$, whereas the increased flame speed reduces $\Upsilon_{\mathrm{f}} = \Upsilon_{\mathrm{f}}^{F} (F>1) > \Upsilon_{\mathrm{f}}^{F,E} (F>1; E>1)$.

Importantly, a single efficiency factor cannot simultaneously restore both $Ka_{\mathrm{f}}$ and $Da_{\mathrm{f}}$ to their unthickened values.
From Eq.~\eqref{eq:Ka_Da_FE}, the condition
$Ka_{\mathrm{f}}^{F,E}=Ka_{\mathrm{f}}$ requires $E=F^{1/3}$, whereas
$Da_{\mathrm{f}}^{F,E}=Da_{\mathrm{f}}$ requires $E=F$.
Except for the trivial case $F=1$, these conditions cannot be satisfied simultaneously.
This is also evident from the premixed combustion-regime diagram, where a given pair of $Ka_{\mathrm{f}}$ and $Da_{\mathrm{f}}$ uniquely defines an operating point through the intersection of the corresponding constant-$Ka_{\mathrm{f}}$ and constant-$Da_{\mathrm{f}}$ lines (see Fig.~\ref{fig:regime_diagram_base}).

For the present analysis, the concept of recovering the flame consumption speed $s_{\mathrm{c}}$ is illustrated by combining the mean local flame-speed model of Howarth et al.~\cite{howarth2023} (Eq.~\eqref{eq:ss_sf_fit_Kaf}) with the SSL flame-wrinkling model (Eq.~\eqref{eq:ssl_lsl_peters}).
For the unthickened flame,
\begin{equation}
\begin{aligned}
    s_{\mathrm{c}}
    &= s_{\mathrm{s}}\Psi
    = s_{\mathrm{f}}\frac{s_{\mathrm{s}}}{s_{\mathrm{f}}}\Psi
    \\
    &\approx s_{\mathrm{f}}
    \bigl(1+0.26e^{-0.038\omega_2}\sqrt{Ka_{\mathrm{f}}}\bigr)
    \\
    &\quad\
    \bigl(1+\gamma_{\mathrm{f}}Ka_{\mathrm{f}}Da_{\mathrm{f}}\bigr).
\end{aligned}
\label{eq:sc_unthickened_efficiency}
\end{equation}
The corrected thickened flame should reproduce this consumption speed, i.e.,
\begin{equation}
    s_{\mathrm{c}}
    \overset{!}{=}
    s_{\mathrm{f}}E
    \left(
        \frac{s_{\mathrm{s}}}{s_{\mathrm{f}}}
    \right)^{F,E}
    \Psi^{F,E}.
\end{equation}
Substitution of Eq.~\eqref{eq:Ka_Da_FE} into the corresponding mean-local-flame-speed and wrinkling models gives

\begin{equation}
\begin{aligned}
&\bigl[1+0.26e^{-0.038\omega_2}\sqrt{Ka_{\mathrm{f}}}\bigr]
\bigl[1+\gamma_{\mathrm{f}}Ka_{\mathrm{f}}Da_{\mathrm{f}}\bigr]
\\
&=E\biggl[
1+0.26e^{-0.038\omega_2}
\sqrt{Ka_{\mathrm{f}}}\frac{F^{1/4}}{E^{3/4}}
\biggr]
\\
&\quad
\biggl[
1+\gamma_{\mathrm{f}}
\frac{Ka_{\mathrm{f}}Da_{\mathrm{f}}}{\sqrt{EF}}
\biggr].
\end{aligned}
\label{eq:efficiency_iterative}
\end{equation}
Equation~\eqref{eq:efficiency_iterative} constitutes an implicit nonlinear condition for $E$.
The efficiency factor must therefore be determined implicitly because changing $E$ modifies the effective $Ka_{\mathrm{f}}^{F,E}$ and $Da_{\mathrm{f}}^{F,E}$, which in turn modifies the mean local flame speed and flame wrinkling that $E$ is intended to correct.

\begin{figure}
    \centering
    \includegraphics[width=\linewidth]{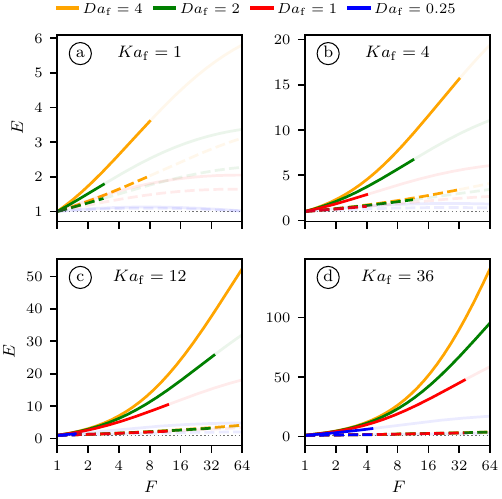}
    \caption{
    Efficiency factor $E$ as a function of the thickening factor $F$ for different prescribed $Ka_{\mathrm{f}}$ and $Da_{\mathrm{f}}$.
    Solid lines represent $E$ obtained from the implicit correction in Eq.~\eqref{eq:efficiency_iterative}, whereas dashed lines show the direct correction, which does not account for the feedback of $E$ into the effective Karlovitz and Damköhler numbers.
    Reduced opacity indicates $\Lambda_{\mathrm{f}}^{F}<1$.
    }
    \label{fig:efficiency_factor_iterative_simple}
\end{figure}

Figure~\ref{fig:efficiency_factor_iterative_simple} shows the resulting efficiency factor as a function of $F$ for different prescribed $Ka_{\mathrm{f}}$ and $Da_{\mathrm{f}}$.
For each parameter combination, the solution is continued from the known value $E=1$ at $F=1$.
At each subsequent value of $F$, the root continuously connected to the preceding solution is determined numerically.
For comparison, a direct correction is also shown.
In this approach, the efficiency factor is obtained from the thickening-induced changes in $Ka_{\mathrm{f}}$ and $Da_{\mathrm{f}}$ before application of $E$, so the feedback of $E$ into $Ka_{\mathrm{f}}^{F,E}$ and $Da_{\mathrm{f}}^{F,E}$ is not accounted for.

The substantial differences between the two approaches demonstrate that this feedback can be significant; their relative difference is shown in the Supplementary Material (Fig.~S.24).
A direct correction based only on the change in consumption speed caused by thickening may therefore be insufficient, because the application of the resulting efficiency factor itself modifies the effective $Ka_{\mathrm{f}}^{F,E}$ and $Da_{\mathrm{f}}^{F,E}$.
Accounting for this coupling through the implicit formulation in Eq.~\eqref{eq:efficiency_iterative} provides a consistent correction within the adopted mean-local-flame-speed and wrinkling models.

Importantly, the need for an implicit correction is not specific to TD unstable lean \ce{H2} flames.
Even for unity-Lewis-number flames, where thickening-induced changes in the mean local flame speed may be negligible, the efficiency factor modifies the effective turbulence--flame interaction parameters and, consequently, flame wrinkling.
This feedback must therefore also be considered when correcting the consumption speed of such flames.

This broader applicability also extends to the choice of models.
Although illustrated here using the mean local flame-speed model of Howarth et al.~\cite{howarth2023} and the SSL wrinkling model, the implicit framework can accommodate other suitable models describing the dependence of these contributions on the relevant turbulence--flame interaction parameters.
The applicability of the correction is therefore not inherently restricted to the low- and moderate-Karlovitz-number range considered by the specific closures employed here.

For the present forced-HIT configurations, $Ka_{\mathrm{f}}$ and $Da_{\mathrm{f}}$ are defined from the prescribed global turbulence conditions.
Accordingly, the analysis above provides a global efficiency factor $E$ for each combination of $F$, $Ka_{\mathrm{f}}$, and $Da_{\mathrm{f}}$.
In a turbulent flame, however, the relevant turbulence and flame parameters vary in space and time.
The same concept could then be applied locally by determining
\begin{equation}
    E = E(x_{i},t)
\end{equation}
from local estimates of $Ka_{\mathrm{f}}$ and $Da_{\mathrm{f}}$.
Such estimates could, for example, be based on turbulent velocity and length scales derived from locally available or transported turbulence quantities, similar to the approach employed in the turbulent efficiency function of Charlette et al.~\cite{charlette2002}.
The resulting spatially and temporally varying efficiency factor would then compensate locally for the effect of artificial thickening on the mean flame propagation and flame wrinkling.

Thus, an on-the-fly solution of the implicit equation is required primarily when $F$, $Ka_{\mathrm{f}}$, and/or $Da_{\mathrm{f}}$ vary locally and/or temporally.
For practical applications, the correction may alternatively be precomputed over the relevant parameter range and represented by tabulated values or fitted functions, thereby avoiding a nonlinear solve at every grid point during runtime.

Future work is required to assess whether the proposed implicit efficiency correction recovers the unthickened flame consumption speed in fully coupled ATF simulations.

\section{Discussion and conclusions}
\label{sc:conclusion}

In this study, flame-resolved simulations with detailed chemistry of a FIAB configuration subjected to HIT are performed to systematically assess the effects of artificial flame thickening on lean turbulent \ce{H2} flames under operating conditions that are susceptible to TD instabilities.
The turbulent conditions are systematically varied in terms of $Ka_{\mathrm{f}}$ and $Da_{\mathrm{f}}$ by varying the velocity fluctuations imposed through the turbulent forcing and the integral length scale through the domain size, while different thickening factors $F$ are considered within the ATF framework.
No efficiency function $E$ is employed in order to isolate and quantify the direct effects of artificial flame thickening on the propagation and structure of TD-unstable turbulent \ce{H2} flames.
These effects are assessed separately at the local and global scales through their influence on the local flame speed and flame wrinkling.

The main conclusions of this study are summarised as follows:
\begin{itemize}[
    itemsep=0pt,
    topsep=0pt,
    parsep=0pt,
    partopsep=0pt
]
\item Artificial flame thickening modifies the effective Karlovitz number experienced by the flame.
Based on the scaling introduced in Sec.~\ref{sec:theory}, increasing the flame thickness by a factor $F$ yields $Ka_{\mathrm{f}}^{F}=Ka_{\mathrm{f}}\sqrt{F}$.
This scaling is directly supported by simulations in which $F$ and the prescribed Karlovitz number $Ka_{\mathrm{f}}$ are varied while $Ka_{\mathrm{f}}^{F}$ and the ratios of the integral and domain length scales to the thickened flame thickness are kept constant.
Despite substantial variations in $F$ and, consequently, in $Ka_{\mathrm{f}}$, only negligible variations in the mean local flame speed $s_{\mathrm{s}}$ are observed at fixed $Ka_{\mathrm{f}}^{F}$.
This demonstrates that $Ka_{\mathrm{f}}^{F}$, rather than the prescribed $Ka_{\mathrm{f}}$, governs the local flame-speed response of the thickened flame.
\item For low-to-moderate Karlovitz numbers, the mean local flame speed is well described by the correlation of Howarth et al.~\cite{howarth2023}, for which the flame-speed enhancement scales approximately with $\sqrt{Ka_{\mathrm{f}}}$.
Combined with $Ka_{\mathrm{f}}^{F}=Ka_{\mathrm{f}}\sqrt{F}$, this implies an approximately $F^{1/4}$ dependence of the local flame-speed enhancement.
Simulations performed at fixed prescribed $Ka_{\mathrm{f}}$ while varying $F$, and therefore $Ka_{\mathrm{f}}^{F}$, confirm this scaling as long as $Ka_{\mathrm{f}}^{F}$ remains within the range of validity of the correlation.
\item Artificial thickening also introduces important length-scale effects in turbulent flames.
At fixed $L_{\mathrm{domain}}$ and $L_{\mathrm{int}}$, increasing $F$ causes the thickened flame thickness $F l_{\mathrm{f}}$ to eventually exceed the integral length scale and, for sufficiently strong thickening, approach relevant geometric scales.
Once $\Lambda_{\mathrm{f}}^{F}=L_{\mathrm{int}}/(F l_{\mathrm{f}})$ falls below unity, the enhancement of the mean local flame speed is progressively suppressed and approaches the one-dimensional laminar-flame limit.
Additional simulations with periodic forcing, allowing $L_{\mathrm{domain}}$ and $L_{\mathrm{int}}$ to be varied independently, demonstrate that increasing the domain size alone does not recover the flame-speed enhancement obtained at larger integral length scales.
Thus, the integral length scale can remain restrictive even without significant geometric confinement.
It should be noted, however, that in this regime the thickened flame thickness lies outside the inertial subrange, such that the assumptions underlying the definition of the Karlovitz number are no longer fulfilled; $Ka_{\mathrm{f}}^{F}$ is no longer a reliable quantitative measure of turbulence–flame interaction.
Additionally, such conditions lie beyond those typically targeted in LES and are therefore primarily relevant to modelling frameworks in which the turbulent scales are fully modelled, such as RANS.
Extending the present framework to RANS-oriented ATF applications will be addressed in future work.
\item The established scaling of flame wrinkling with the effective Damköhler number $Da_{\mathrm{f}}^{F}=Da_{\mathrm{f}}/F$ also holds for the TD unstable lean hydrogen flames considered in the present work.
When the prescribed $Da_{\mathrm{f}}$ is used, substantial scatter is observed, whereas accounting for thickening through $Da_{\mathrm{f}}^{F}$ results in a pronounced collapse of the data.
The resulting trends are well described by the SSL wrinkling model originating from Damköhler~\cite{damkoehler1940}, using model coefficients adopted from~\cite{hunt2025_phd}.
Consistently, at fixed prescribed $Ka_{\mathrm{f}}$ and $Da_{\mathrm{f}}$, the normalised wrinkling enhancement $(\Psi-1)$ exhibits an approximately $F^{-1/2}$ dependence, in agreement with the scaling implied by the SSL.
\item Artificial thickening therefore affects the two contributions to the global flame propagation in opposite directions.
The increase in $Ka_{\mathrm{f}}^{F}$ enhances the mean local flame speed in the regime investigated here, whereas the decrease in $Da_{\mathrm{f}}^{F}$ reduces flame wrinkling.
These contributions exhibit different dependencies on $F$, such that the net flame consumption speed generally decreases with increasing $F$ relative to an unthickened reference flame at the same prescribed $Ka_{\mathrm{f}}$ and $Da_{\mathrm{f}}$, provided that $\Lambda_{\mathrm{f}}^{F}\geq1$.
These findings have direct implications for ATF modelling:
Recovering the consumption speed of the corresponding unthickened flame requires compensating for both effects introduced by artificial thickening: the modification of the local flame speed through $Ka_{\mathrm{f}}^{F}$ and the modification of flame wrinkling through $Da_{\mathrm{f}}^{F}$.
A correction or efficiency function should therefore account for both contributions rather than correcting flame wrinkling alone.
\item Importantly, the present results suggest that such corrections do not necessarily require models specifically developed for artificially thickened flames.
Instead, sufficiently accurate models for the local flame speed and flame wrinkling over the relevant ranges of Karlovitz and Damköhler numbers can, in principle, be applied to both the unthickened and thickened flame states.
The effect of artificial thickening then enters through the modified flame thickness and, consequently, through the resulting $Ka_{\mathrm{f}}^{F}$ and $Da_{\mathrm{f}}^{F}$.
The corresponding model predictions can be used to construct correction factors that compensate for the difference between the thickened and intended flame states.
However, because an efficiency factor applied to the reaction source terms also modifies the effective flame speed and thereby feeds back on the corresponding Karlovitz and Damköhler numbers ($Ka_{\mathrm{f}}^{F,E}$ and $Da_{\mathrm{f}}^{F,E}$), the correction is inherently coupled.
An implicit formulation for the efficiency factor is therefore proposed that accounts for this feedback.
\item Even if the global flame consumption speed is accurately recovered through appropriate modelling corrections, the inherent modifications of the effective $Ka_{\mathrm{f}}^{F,E}$ and $Da_{\mathrm{f}}^{F,E}$, together with the changes in the characteristic length scales of flame-front structures, can lead to substantial differences in flame morphology.
This is particularly relevant for TD-unstable flames, where TD effects additionally influence the flame-front structure and the spatial distribution of local flame properties, such as propagation speed and thickness.
Consequently, accurately reproducing the global consumption speed does not necessarily imply that the local flame structure resembles that of the corresponding unthickened reference flame.

\end{itemize}
The next step is to assess this correction framework \textit{a posteriori} in fully coupled LES of turbulent flames.
Both global efficiency functions based on characteristic Karlovitz and Damköhler numbers and local formulations based on local estimates of $Ka_{\mathrm{f}}$ and $Da_{\mathrm{f}}$, together with the local thickening factor $F$, should be investigated.
Such simulations, including more complex configurations such as turbulent jet flames, would enable evaluation of the proposed correction under spatially and temporally varying turbulence conditions.
Furthermore, the mean local flame-speed model of Howarth et al.~\cite{howarth2023} should be extended and validated at higher Karlovitz numbers, as artificial thickening increases the effective Karlovitz number, potentially exceeding the range covered by existing correlations even at moderate values of the prescribed $Ka_{\mathrm{f}}$.

\section*{CRediT authorship contribution statement}
\textbf{M. Schneider}: Conceptualization, Methodology, Investigation, Software, Formal analysis, Data Curation, Visualization, Writing - original draft.
\textbf{V. Schuh}: Conceptualization, Methodology, Investigation, Software, Formal analysis, Visualization, Writing - review \& editing.
\textbf{F. Rong}: Investigation, Visualization, Writing - review \& editing.
\textbf{D. Kaddar}: Investigation, Formal analysis, Data Curation, Visualization, Writing - review \& editing.
\textbf{A. J. Aspden}: Methodology, Formal analysis, Investigation, Supervision, Writing - review \& editing.
\textbf{T. L. Howarth}: Methodology, Formal analysis, Investigation, Software, Writing - review \& editing.
\textbf{C. Hasse}: Conceptualization, Resources, Supervision, Project administration, Funding acquisition, Writing - review \& editing.
\textbf{H. Nicolai}: Conceptualization, Methodology, Formal analysis, Supervision, Project administration, Writing - review \& editing.

 \section*{Declaration of competing interest}
The authors declare that they have no known competing financial interests or personal relationships that could have appeared to influence the work reported in this paper.

\section*{Declaration of generative AI and AI-assisted technologies in the manuscript preparation process}
During the preparation of this work, the authors used ChatGPT (OpenAI) to improve the language, clarity, and readability of the manuscript.
All AI-assisted suggestions were critically reviewed and edited by the authors, who take full responsibility for the content of the publication.

\section*{Acknowledgments}
This work was funded by the Deutsche Forschungsgemeinschaft (DFG, German Research Foundation) under project number 523792378 within the framework of Priority Programme SPP 2419.
FR acknowledges additional funding from the DFG under project number 310695286.
DK acknowledges additional funding from the JUPITER AI Factory (JAIF) under grant agreement number 101250682.
HN acknowledges funding from the Natural Sciences and Engineering Research Council of Canada (NSERC) under grant number RGPIN-2026-04601.
The authors gratefully acknowledge the Gauss Centre for Supercomputing e.V. (GCS) for providing computing time on the supercomputers JUWELS and JUPITER at the Jülich Supercomputing Centre (JSC) through the projects efficient and h2-max.

\FloatBarrier

\bibliographystyle{elsarticle-num-title-doi}
\bibliography{bibliography}

\end{document}